\documentclass[11pt,a4paper,preprintnumbers,nofootinbib, superscriptaddress]{revtex4}

\usepackage[T1]{fontenc}
\usepackage[utf8]{inputenc}
\usepackage{amsmath}
\usepackage{amssymb}
\usepackage{hyperref}
\usepackage[final]{graphicx}

\usepackage[english]{babel}

\usepackage{graphicx}
\usepackage[dvipsnames]{xcolor}
\usepackage{subfigure}

\usepackage{braket}

\usepackage{amsfonts}
\usepackage{amstext}
\usepackage{slashed}

\usepackage{microtype}
\usepackage{scalerel}

\usepackage{multirow}
\usepackage{booktabs}

\usepackage{placeins}
\usepackage{verbatim}

\DeclareMathOperator{\Tr}{Tr}

\newcommand{\im}{\mathrm{i}}

\def\GeV{{\rm GeV}}

\def\A{\mathcal{A}}
\def\B{\mathcal{B}}

\def\L{\mathcal{L}}

\def\O{\mathcal{O}}

\def\ubar{\overline{u}}
\def\dbar{\overline{d}}
\def\sbar{\overline{s}}

\def\qbar{\overline{q}}

\begin{document}

\title{Rare radiative decays of charm baryons}

\author{Nico Adolph}
\email{nico.adolph@tu-dortmund.de}
\author{Gudrun Hiller}
\email{gudrun.hiller@tu-dortmund.de}
\affiliation{TU Dortmund University, Department of Physics, Otto-Hahn-Str.4, D-44221 Dortmund, Germany}
\preprint{DO-TH 21/26}

\begin{abstract}
 We study weak radiative $|\Delta c|=|\Delta u|=1$ decays of the charmed anti-triplett ($\Lambda_c$, $\Xi_c^{+}$, $\Xi_c^{0}$)
 and sextet ($\Sigma_c^{++}$, $\Sigma_c^+$, $\Sigma_c^0$, $\Xi_c^{\prime +}$, $\Xi_c^{\prime 0}$, $\Omega_c$) baryons
 in the standard model (SM) and beyond. We work out $SU(2)$ and $SU(3)_F$-symmetry relations. We propose to study self-analyzing  decay chains
  such as $\Xi_c^+ \to \Sigma^+ (\to p \pi^0) \gamma$ and  $\Xi_c^0 \to \Lambda (\to p \pi^-) \gamma$,
 which  enable new physics sensitive polarization studies.  SM contributions can be controlled by corresponding analysis of the Cabibbo-favored decays
 $\Lambda_c^+ \to \Sigma^+ (\to p \pi^0) \gamma$ and  $\Xi_c^0 \to \Xi^0 (\to \Lambda \pi^0) \gamma$.
 Further tests of the SM are available with initially polarized  baryons including $\Lambda_c \to p 
 \gamma$ together with $\Lambda_c \to \Sigma^+ \gamma$ decays, or 
 $\Omega_c \to \Xi^0 \gamma$  together with $\Omega_c \to (\Lambda,\Sigma^0) \gamma$.
In addition,
 CP-violating new physics contributions to dipole operators can enhance CP-asymmetries up to few percent.
\end{abstract}

\maketitle

\section{Introduction}

Rare decays of charmed hadrons are sensitive to flavor in and beyond the standard model (SM). The severe Glashow-Iliopoulos-Maiani (GIM) suppression of $|\Delta c|=|\Delta u|=1$ couplings within 
the SM makes observable electroweak effects  of new physics (NP) origin. At the same time, resonance contributions and limited control in heavy quark methods at the charm mass
prohibit to probe short-distance physics in simple observables such as branching ratios. Null tests are therefore key in testing the SM in rare charm decays, in addition
to data-driven methods to control the SM background. 
Despite its nominal uncertainties of order 30 percent, U-spin and $SU(3)_F$-symmetries are useful in cases such as radiative modes
where kinematical cuts, or angular distributions are not available, while NP effects can be huge.
In fact, the presence of partner modes,  one induced at  tree level in SM via $\bar u c \bar s d$  or $\bar u c \bar d s$
and one subject to $\bar u c \bar q q$, $q=u,d,s$, sensitive to NP, makes $SU(3)_F$-analyses in charm more powerful than in beauty, where no such partner decays exist.
To stress this point even some more: while flavor symmetries connect rare $b$-decays~\cite{Gronau:2000zy}, unlike in $c \to u$ transitions, there is no link to 
a SM-dominated ($W$-induced) mode, that would allow to determine experimentally the SM background.
Previous works exploiting a data-driven strategy with rare radiative charm decays and partner modes proposed time-dependent analysis of $D,\bar D \to V \gamma$, $V=\rho, K^*, \phi$, \cite{deBoer:2018zhz} and top-down asymmetries  
in $D \to K_1 \gamma$ \cite{deBoer:2018zhz,Adolph:2018hde}.

Here we analyze the NP potential of radiative charm baryon decays. Previous works exist for polarized $\Lambda_c \to p \gamma$ decays  \cite{deBoer:2017que}.
We propose to measure  the photon polarization that can be studied
using self-analyzing secondary decays. Hyperons with sizable branching ratios and weak decay parameter \cite{Zyla:2020zbs} are given in 
Table \ref{tab:hyperons}.
Suitable singly- Cabibbo suppressed (SCS) decay chains turn out to be
$\Xi_c^+ \to \Sigma^+ (\to p \pi^0) \gamma$ and $\Xi_c^0 \to \Lambda (\to p \pi^-) \gamma$  with Cabibbo-favored  (CF) partners $\Lambda_c^+ \to \Sigma^+ (\to p \pi^0) \gamma$ and  $\Xi_c^0 \to \Xi^0 (\to \Lambda \pi^0) \gamma$, respectively.

Branching ratio estimates are subject to sizable  uncertainites, see  \cite{Singer:1996ba} for  SCS modes.
Theory predictions for CF  modes differ significantly  \cite{Uppal:1993, Cheng:1994kp, Kamal:1983}, with  branching ratios at the level of  $10^{-4}$,
consistent with hierarchies from the SM weak annihilation mechanism in $D$-meson decays \cite{deBoer:2017que}.
However, these uncertainties do not affect the strategy to test the SM; we only need them here to estimate the NP reach. Ultimately, the branching ratios have to be determined
by experiment.
 \begin{table}[t]
  \centering\begin{tabular}{c|c|c}
    Decay & $\B$ & $\alpha_B$ \\
    \hline
    $\Lambda(1116)\to p \pi^-$ & $(63.9 \pm 0.5) \%$ & $0.732 \pm 0.014$ \\
     $\Sigma^+(1189) \to p \pi^0$ & $(51.57 \pm 0.30) \%$ & $-0.982 \pm 0.014$ \\
      $\Xi^0(1315) \to \Lambda \pi^0$ & $(99.52 \pm 0.012)\%$ & $-0.356 \pm 0.011$ \\
  \end{tabular}
  \caption{Branching ratio  $\B$ and weak decay parameter $\alpha_B$ of self-analyzing hyperon decays   \cite{Zyla:2020zbs}.
  Note, the $\Xi^-(1322)$ decays almost entirely to $\Lambda \pi^-$ with sizable $\alpha_B=-0.4$, however, it  is not produced in rare decays of charm baryons.} 
  \label{tab:hyperons}
\end{table}
Rare radiative charm baryon decays can be studied at high luminosity flavor facilities, such as LHCb~\cite{Cerri:2018ypt}, Belle II~\cite{Kou:2018nap}, BES III~\cite{Ablikim:2019hff}, and possible future machines~\cite{Charm-TauFactory:2013cnj,Abada:2019lih}. We stress that none of the rare radiative charm baryon modes has been observed yet. 

The paper is  organized  as follows:
In Sec.~\ref{sec:th} we give  the effective Lagrangian \cite{deBoer:2017que}, and present the $SU(3)_F$ and $U$-spin decompositions for
anti-triplet to octet and sextet to octet decays. We also obtain sum rules and NP sensitivity ratios.
In Sec.~\ref{sec:obs} we present observables for two-body, and self-analyzing three-body decays, relate branching ratios and work out CP-asymmetries.
We work out the BSM reach in Sec.~\ref{sec:BSM reach}.
We conclude in Sec.~\ref{sec:con}.
In App.~\ref{app:parameter} we give parametric input to our numerical analysis.  $SU(2)$-flavor decompositions are given in App.\ref{app:flavor symmetrien}.
The $SU(3)$-decomposition of the decay amplitudes  provided in App.~\ref{sec:SU(3) decomposition}.
Relations for  sextet to decuplet  decays are deferred to App.~\ref{app:6to10}.
Irreducible $SU(3)_F$ amplitudes are given in App.~\ref{app:SU3}.
In App.~\ref{app:form_factors} hadronic  tensor form factors are defined.

\section{Theoretical description of radiative charm baryon decays \label{sec:th}}

\subsection{Effective weak Lagrangian}

We use the framework of weak effective  theory in which the SCS, CF and DCS decay Lagrangians can be written in terms of  dimension six operators \cite{deBoer:2017que}
\begin{align} \label{eq:eff}
    \L_{\text{eff}}^{\rm SCS} = \frac{4G_F}{\sqrt{2}}\left(\sum_{q=d,s}V_{cq}^* V_{uq}\sum_{i=1}^2 C_i O_i^{(q q)} + \sum_{i=3}^6 C_i O_i + \sum_{i=7}^8 \left(C_i O_i + C_i^\prime O_i^\prime\right)\right)\, ,
\end{align}
\begin{align}
    \L_{\text{eff}}^{\rm CF} = \frac{4G_F}{\sqrt{2}} V_{cs}^* V_{ud}\sum_{i=1}^2 C_i O_i^{(ds)} \, , \quad 
    \L_{\text{eff}}^{\rm DCS} = \frac{4G_F}{\sqrt{2}} V_{cd}^* V_{us}\sum_{i=1}^2 C_i O_i^{(sd)}\, ,
\end{align}
where $G_F$ is Fermi's constant and $V_{ij}$ are elements of the Cabibbo-Kobayashi-Maskawa (CKM) matrix. In the following, we only consider the  SM contributions of the four quark operators $O_{1,2}^{(q q^\prime)}$ and  BSM effects to the electromagnetic dipole operators $O_7^{(\prime)}$, $q,q^\prime=d,s$,
\begin{equation}
  \begin{alignedat}{2}
    &O_1^{(q q^\prime )} = \left(\ubar_L \gamma_\mu T^a q_L\right) \left(\qbar_L^\prime \gamma^\mu T^a c_L\right)\, , \quad &&O_2^{(q q^\prime)} = \left(\ubar_L \gamma_\mu q_L\right) \left(\qbar_L^\prime \gamma^\mu c_L\right)\, ,\\
    &O_7 = \frac{e m_c}{16\pi^2}  \left(\ubar_L \sigma^{\mu \nu} c_R\right)F_{\mu \nu}\, , &&O_7^\prime = \frac{e m_c}{16\pi^2}  \left(\ubar_R \sigma^{\mu \nu} c_L\right)F_{\mu \nu}\,.
  \end{alignedat}
\end{equation}
Here, $q_{L/R}$ are chiral quark fields, $T^a$ are the generators of $SU(3)$ normalized to $\Tr(T^a T^b) = \delta^{ab}/2$, $\sigma^{\mu \nu}= \frac{i}{2}\left[\gamma^\mu, \gamma^\nu\right]$ and $F_{\mu \nu}$ is the photon field strength tensor.  Due to an efficient GIM cancellation and resulting small SM Wilson coefficients, we safely neglect the effects of the QCD penguin operators $O_{3-6}$, as well as  the chromomagnetic dipole operators $O_8^{(\prime)}$, which enter at higher orders in radiative decays.
At the charm scale $\mu_c \in \left[m_c/\sqrt{2}, \sqrt{2}m_c\right]$, the leading order Wilson coefficients of the four-quark operators are given by \cite{deBoer:2017que}
\begin{equation} \label{eq:c4f}
  \begin{alignedat}{2}
    &C_1 \in [-1.28, -0.83]\, , \qquad &&C_2 \in [1.14, 1.06]\, ,\\
    &C_+ = C_2 + \frac{1}{3}C_1 \in \left[0.76, 0.78\right] \, , \qquad &&C_- = C_2 - \frac{2}{3}C_1 \in \left[1.99, 1.61\right]\, ,\\
    & \tilde{C} = \frac{4}{9}C_1 + \frac{1}{3}C_2 \in \left[-0.189, -0.018\right]\, . &&
  \end{alignedat}
\end{equation}
Here, we use $m_c=1.27\, \GeV$. All other coefficients in (\ref{eq:eff}) are severly GIM-suppressed, and negligible for phenomenology in the SM.
Specifically, the effective coefficient of the dipole operator  $C_7^\text{eff} $ is of order $ \O(10^{-3})$ \cite{deBoer:2017que}.

 Physics beyond the SM can significantly increase the Wilson coefficients $C_7^{(\prime)}$. $D \to \rho^0 \gamma$ and $D \to \pi \ell \ell$ decays yield the model independent constraints \cite{deBoer:2018buv, Abdesselam:2016yvr, Bause:2019vpr}
\begin{align} \label{eq:c7range}
  |C_7|, |C_7^\prime| \lesssim 0.3\, .
\end{align}

\subsection{Decay amplitudes}

The general Lorentz decomposition of the $B_c(P, s_{B_c}) \to B(q, s_B) \gamma(k, \epsilon^*)$ amplitude is given by 
\begin{align} \label{eq:amp}
  \A(B_c \to B \gamma) = \frac{G_F e}{\sqrt{2}} \ubar(q, s_B) \left[ F_L P_R+F_R P_L \right] \slashed{k} \slashed{\epsilon}^* u(P, s_{B_c}) 
 \end{align}
where $P_L=(1- \gamma_5)/2$, $P_R=(1+ \gamma_5)/2$ are chiral projectors, and $F_L$ and $F_R$ denote the contributions for left-handed and right-handed photons, respectively. Here,
$s_{B_c}$ ($s_B)$ denotes the spin of the $B_c$ ($B$) baryon, and $P,q,k$ refer to the four-momenta of the $ B_c$, $B$ and photon, respectively.

\begin{figure}[t]
  \centering
  \includegraphics[width=0.8\linewidth]{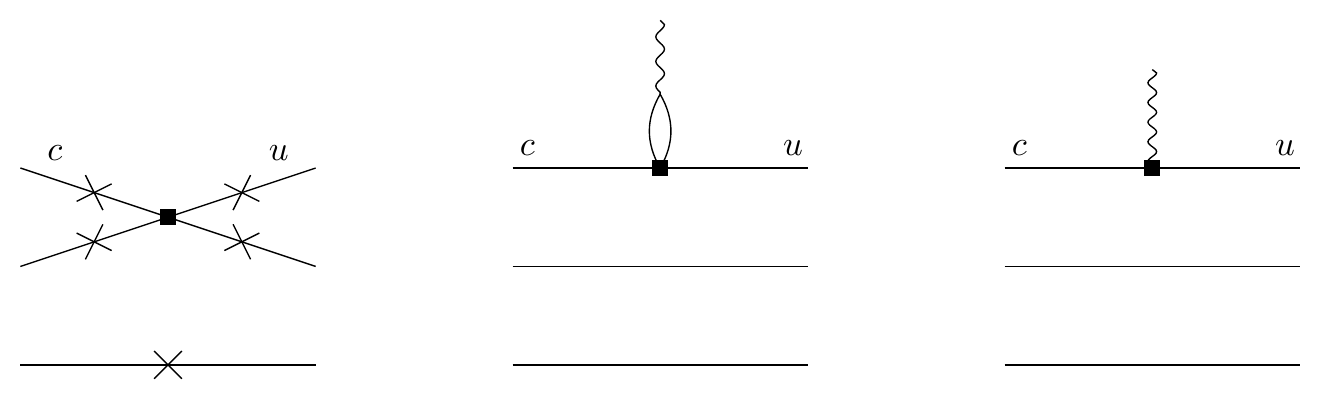}
  \caption{Contributions to $B_c \to B \gamma$ decays. Weak annihilation inside the baryon is shown on the left. The crosses denote the possible photon emissions. Long distance contributions with a photon coupling to the weak current via the light vector mesons $\rho^0$, $\omega$ and $\phi$ are illustrated  in the middle. On the right the contributions from the electromagnetic dipole operators $O_7^{(\prime)}$ are shown. The latter are negligible within the SM, but can induce sizable contributions beyond the SM.}
  \label{fig:diagrams}
\end{figure}
Contributions to  $B_c \to B \gamma$ decays from different mechanisms are illustrated in 
Fig.~\ref{fig:diagrams}. The weak annihilation (WA) diagrams on the left provide the dominant contributions to the SM amplitude; at leading order they scale with the color-allowed  coefficent $C_-$.  We do not attempt to compute the WA amplitude as available theory methods lack sufficient control.  Instead we propose to extract them from SM-dominated CF (or DCS) decays once measured and use them for the SCS modes to test the SM using flavor symmetries.

The contributions via intermediate vector resonances shown in Fig.~\ref{fig:diagrams} (middle) depends on the color-suppressed combination $\tilde{C}$ of Wilson coefficients. 
To obtain a gauge invariant amplitude for the long distance contributions \cite{Singer:1996ba}, the vector mesons have to be transversal polarized. This can be achieved by the Golowich-Pakvasa method \cite{Deshpande:1996, Golowich:1995}. The  long distance amplitudes read 
\begin{align}
  \begin{split}
    &F_L^{\text{LD}} = - 2\frac{\tilde{C} C_{\text{VMD}}}{m_c} h_\perp^{B_c \to B}(k^2=0)\,, \qquad F_R^{\text{LD}} = 0\, ,
  \end{split}
\end{align}
where
\begin{align}
  C_{\text{VMD}}=V_{cs}^* V_{us} \left(-\frac{1}{3}f_\phi^2\right) + V_{cd}^* V_{ud}\left(-\frac{1}{2}{f^{(d)}_{\rho^0}}^2 +\frac{1}{6}{f^{(d)}_\omega}^2\right)\,.
\end{align}
Using $f_V(k^2=0) \approx f_V(k^2=m_V^2)$ one obtains $C_{\text{VMD}} \sim  - 6.3  \cdot 10^{-4}\, \GeV^2$,  due to GIM cancellations \cite{Singer:1996ba} about one to two orders of magnitude smaller than individual contributions $ \lambda f_V^2 \sim 10^{-2}\, \GeV^2$.
With the color and GIM  suppression these long distance contributions are negligible compared to the weak annihilation contributions.

Contributions from  electromagnetic dipole operators $O_7^{(\prime)}$, shown in Fig.~\ref{fig:diagrams} in the diagram on the right, read
\begin{align} \label{eq:NP}
  \begin{split}
    &F_L^\text{NP} = -\frac{m_c}{2\pi^2} C_7 h_\perp^{B_c \to B}(k^2=0)\,,\\
    &F_R^\text{NP} = -\frac{m_c}{2\pi^2} C_7^\prime h_\perp^{B_c \to B}(k^2=0)\,,
  \end{split}
\end{align}
which can be neglected within the SM due to an efficient GIM cancellation. 
We employ (\ref{eq:NP}) to estimate the NP reach. Hadronic transition form factors for dipole currents $h_\perp$ and $\tilde{h}_\perp$  are defined in App.~\ref{app:form_factors}.
Furthermore, we used the endpoint relation $h_\perp^{B_c \to B}(k^2=0) = \tilde{h}_\perp^{B_c \to B}(k^2=0)$ of the tensor form factors \cite{Hiller:2021zth}.  For $\Lambda_c \to p$ they are known from lattice QCD \cite{Meinel:2017ggx} and relativistic quark models \cite{Faustov:2018dkn}. Moreover, results from Light cone sum rules are available for $\Xi_c \to \Sigma$ \cite{Azizi:2011mw}. In our numerical analysis, we use the results from lattice QCD $h_\perp^{\Lambda_c \to p}(k^2=0)= 0.511 \pm 0.027$ and Iso-/U-spin relations  between different baryonic transitions within the same multiplets, derived in Sec.~\ref{sec:flavor}.

\subsection{Flavor symmetry relations}
\label{sec:flavor}

To bypass the difficulties in the calculation of the non-factorizable weak annihilation amplitude, it can be estimated using data on branching ratios and the photon polarizations for the SM-like decay channels and approximate flavor relations. For this purpose, we express the SM Lagrangian in terms of U-spin operators following \cite{Brod:2012ud}
\begin{align}
  \begin{split}
    \L_{\text{eff}}^{\text{CF}} &\propto -V_{cs}^* V_{ud} (1,-1)_U \,,\\
    \L_{\text{eff}}^{\text{SCS}} &\propto  \sqrt{2} \left( \Sigma (1,0)_U + \Delta (0,0)_U \right)\, , \\
    \L_{\text{eff}}^{\text{DCS}} &\propto  V_{cd}^* V_{us} (1,1)_U\,,\\
  \end{split}
  \label{eq:Leff_SU(2)_Operatoren}
\end{align}
where $(i, j)_X = \O^{X = i}_{X_3 = j}$ and 
\begin{equation}
  \Sigma = \frac{V_{cs}^* V_{us} - V_{cd}^* V_{ud}}{2}\, , \qquad \Delta = \frac{V_{cs}^* V_{us} + V_{cd}^* V_{ud}}{2} = - \frac{V_{cb}^* V_{ub}}{2}\, .
\end{equation}
Due to the CKM suppression of $\lambda^5$, where $\lambda = 0.225$ is the Wolfenstein parameter, the contributions of the singlet operators are negligible for branching ratios and the photon polarization. However, they are crucial for SM CP-asymmetries, see Sec.~\ref{sec:CP}.\\
The general U-spin decomposition of the SM decay amplitudes are given in Table \ref{tbl:U-Spin} and \ref{tbl:U-Spin sextet}. The middle and right diagram in Fig.~\ref{fig:diagrams} show the long distance and short distance $c \to u \gamma$ contributions, respectively, which are only possible for the BSM sensitive SCS modes. 
In terms of U-spin and isospin, the corresponding  Lagrangian can be written as
\begin{align}
  \begin{split}
    \L_\text{eff}^{c \to u \gamma} &\propto (0,0)_U\\
    &\propto(1/2,1/2)_I\, .
  \end{split}
\end{align}
which is useful to derive form factor relations.
In Table \ref{tbl:Iso- U-spin BSM} and \ref{tbl:Iso- U-spin BSM sextet} we show the decomposition of the $c \to u \gamma$ amplitudes.

Similarly, we can use the $SU(3)_F$ symmetry and write the SM Lagrangian as
\begin{align}
  \begin{split}
    \L_\text{eff}^\text{CF} &= V_{cs}^* V_{ud} \left( \overline{\textbf{6}}_{-\frac{2}{3}, 1, 1} + \textbf{15}_{-\frac{2}{3}, 1, 1} \right)\, , \\
    \L_\text{eff}^\text{SCS} &= \Sigma \left( \sqrt{2} ~\overline{\textbf{6}}_{\frac{1}{3}, \frac{1}{2}, \frac{1}{2}} + \frac{2}{\sqrt{3}}\textbf{15}_{\frac{1}{3}, \frac{1}{2}, \frac{1}{2}} - \sqrt{\frac{2}{3}}\textbf{15}_{\frac{1}{3}, \frac{3}{2}, \frac{1}{2}} \right) \\
    &+ \Delta \left(\textbf{3}_{\frac{1}{3}, \frac{1}{2}, \frac{1}{2}} + \frac{1}{\sqrt{3}}\textbf{15}_{\frac{1}{3}, \frac{1}{2}, \frac{1}{2}} + \sqrt{\frac{2}{3}}\textbf{15}_{\frac{1}{3}, \frac{3}{2}, \frac{1}{2}}\right)\, ,\\
    \L_\text{eff}^\text{DCS} &= V_{cd}^* V_{us} \left( -\overline{\textbf{6}}_{\frac{4}{3}, 0, 0} + \textbf{15}_{\frac{4}{3}, 1, 0} \right)\, ,
  \end{split}
\end{align}
where $\textbf{R}_{Y, I, I_3}$ denotes a $SU(3)_F$ operator with irreducible representation $\textbf{R}$, hypercharge $Y$ and isospin $I,I_3$. Note that the operators $\overline{\textbf{6}}$ and $\textbf{15}$ scale with $C_-$ and $C_+$ \cite{Geng:2018plk}, respectively. Furthermore, the triplet operator $\textbf{3}_{\frac{1}{3}, \frac{1}{2}, \frac{1}{2}}$ refers to the long distance contribution shown in the middle of Fig.~\ref{fig:diagrams}. The short distance $c \to u \gamma$ contributions are described by an operator with the same quantum numbers. The $SU(3)_F$-decompositions of the decay amplitudes are shown in Table \ref{tbl:SU(3) triplet} and \ref{tbl:SU(3) sextet}.

In Table \ref{tbl:Flavor Relationen} and \ref{tbl:Flavor Relationen sextet}, we summarize our results for the $SU(2)_U$ and $SU(3)_F$ relations between decay amplitudes and compare them with the $SU(3)_F$ irreducible representation approach (IRA). More information on the $SU(3)_F$ IRA can be found in App.~\ref{app:SU3}. Note that in this work we will focus on decays into the light octet baryons. Thus, the $\Sigma_c^{++}$, which decays exclusively into $\Delta^{++}\gamma$, is not present in Table \ref{tbl:Flavor Relationen sextet}. For completeness, we have added the relations for decays of charmed sextet baryons into decuplet baryons in 
App.~\ref{app:6to10}.
\begin{table}[t]
  \centering\begin{tabular}{l|c|c|c}
    Decay  & U-Spin & $SU(3)_F$ & $SU(3)_F$ IRA\\
    \hline
    $\Lambda_c \to \Sigma^+ \gamma$ & $V_{cs}^* V_{ud}A_\Sigma$         & $ V_{cs}^* V_{ud} B_\Sigma$ & $ V_{cs}^* V_{ud} D$\\
    $\Xi_c^0 \to \Xi^0 \gamma$      & $V_{cs}^* V_{ud}A_\Sigma^\prime $ & $ V_{cs}^* V_{ud} B_\Sigma^\prime$ & $ V_{cs}^* V_{ud} D^\prime$\\
    \hline
    $\Lambda_c \to p \gamma$      & $-\Sigma A_\Sigma + \Delta A_\Delta + A_7$ & $\Sigma B_\Sigma - \Delta B_\Delta + B_7$ & $ \Sigma D - \Delta \tilde{b}_4 + D_7$\\
    $\Xi_c^+ \to \Sigma^+ \gamma$ & $\Sigma A_\Sigma + \Delta A_\Delta + A_7$  & $-\Sigma B_\Sigma - \Delta B_\Delta + B_7$  & $\Sigma D + \Delta \tilde{b}_4 - D_7$\\
    $\Xi_c^0 \to \Lambda \gamma$  & $-\sqrt{\frac{3}{2}} \Sigma A_\Sigma^\prime - \frac{1}{2} (\Delta A_\Delta^\prime  + A_7^{\prime})$ & $\sqrt{\frac{3}{2}} \Sigma B_\Sigma^\prime + \sqrt{\frac{3}{2}} \Delta B_\Delta + \frac{1}{\sqrt{6}} B_7$ & $-\sqrt{\frac{3}{2}}\Sigma D^\prime + \sqrt{\frac{3}{2}} \Delta \tilde{b}_4 + \frac{1}{\sqrt{6}} D_7$\\
    $\Xi_c^0 \to \Sigma^0 \gamma$ & $-\frac{1}{\sqrt{2}} \Sigma A_\Sigma^\prime + \frac{\sqrt{3}}{2} (\Delta A_\Delta^\prime + A_7^\prime)$ & $-\frac{1}{\sqrt{2}} \Sigma B_\Sigma^\prime + \frac{3}{\sqrt{2}} \Delta B_\Delta +  \sqrt{\frac{1}{2}}B_7$ & $\frac{1}{\sqrt{2}}\Sigma D^\prime + \frac{3}{\sqrt{2}} \Delta \tilde{b}_4 +  \frac{1}{\sqrt{2}}D_7$\\
    \hline
    $\Xi_c^+ \to p \gamma$ & $V_{cd}^* V_{us}A_\Sigma$ & $V_{cd}^* V_{us}B_\Sigma$ & $ V_{cd}^* V_{us} D$\\
    $\Xi_c^0 \to n \gamma$ & $-V_{cd}^* V_{us}A_\Sigma^\prime $ & $V_{cd}^* V_{us} B_\Sigma^\prime $& $- V_{cd}^* V_{us} D^\prime$
  \end{tabular}
  \caption{Flavor symmetry relations of the decay amplitudes for the charmed anti-triplet baryons. $A^{(\prime)}_\Sigma$ and $A^{(\prime)}_\Delta$ refer to the U-spin triplet and singlet SM contributions of the W-exchange diagrams. $A_7^{(\prime)} = A_\text{NP}^{(\prime)} + A_\text{LD}^{(\prime)}$ denote the $c \rightarrow u \gamma$ short distance and long distance contributions with intermediate vector resonances. Note that $A_7^{(\prime)}$ originate from U-spin singlet operators and $A_\text{LD}^{(\prime)}$ also scales with $\Delta$ in the limit $f_\phi = f_\rho^{(d)} = f_\omega^{(d)}$. Furthermore, $A_7 = \sqrt{\frac{2}{3}}A_7^{\prime}$ in the isospin limit. An analogous notation is used for the $SU(3)_F$ relations with $B_\Sigma = \sqrt{\frac{2}{3}}A_{\overline{6}}- \frac{2}{\sqrt{15}}A_{15}$, $B^\prime_\Sigma = \sqrt{\frac{2}{3}}A_{\overline{6}}+ \frac{2}{\sqrt{15}}A_{15}$, $B_\Delta = \frac{1}{\sqrt{15}}A_{15}$ and $B_7 = B_\text{NP} + \Delta A_3$. Similarly, $D = -2 (\tilde{b}_1 - \tilde{b}_3 + \tilde{b}_4)$, $D^\prime = 2 (\tilde{b}_1 - \tilde{b}_3 - \tilde{b}_4^\prime)$ and $D_7 = 2b_1 + b_2$ denote the W exchange and $c \to u \gamma$ contributions, respectively. The $b_i, \tilde b_i$ are defined in App.~\ref{app:SU3}.}
  \label{tbl:Flavor Relationen}
\end{table}
\begin{table}[t]
  \centering
  \resizebox{\textwidth}{!}{%
  \begin{tabular}{l|c|c|c}
    Decay  & U-Spin & $SU(3)_F$ & $SU(3)_F$ IRA\\
    \hline
    $\Sigma_c^+ \to \Sigma^+ \gamma$  & $V_{cs}^* V_{ud} E_\Sigma$& $V_{cs}^* V_{ud} F_\Sigma$ & $V_{cs}^* V_{ud} G$\\
    $\Sigma_c^0 \to \Lambda \gamma$  & $V_{cs}^* V_{ud} \left(\frac{\sqrt{3}}{2\sqrt{2}}E_\Sigma^\prime + \frac{1}{2\sqrt{3}}E_\Sigma^{\prime\prime}\right)$& $-\frac{1}{\sqrt{3}}V_{cs}^* V_{ud} (2F_\Sigma^\prime - F_\Sigma)$ & $\frac{1}{\sqrt{3}}V_{cs}^* V_{ud} (2G^\prime - G)$\\
    $\Sigma_c^0 \to \Sigma^0 \gamma$  & $V_{cs}^* V_{ud} \left(\frac{1}{2\sqrt{2}}E_\Sigma^\prime - \frac{1}{2}E_\Sigma^{\prime\prime}\right)$ &$V_{cs}^* V_{ud} F_\Sigma$ & $-V_{cs}^* V_{ud} G$\\
    $\Xi_c^{\prime 0} \to \Xi^0 \gamma$  & $V_{cs}^* V_{ud} \frac{1}{\sqrt{2}}E_\Sigma^\prime$ & $V_{cs}^* V_{ud} F_\Sigma^\prime$ & $V_{cs}^* V_{ud} G^\prime$\\
    \hline
    $\Sigma_c^+ \to p \gamma$  & $-\Sigma E_\Sigma + \Delta E_\Delta + E_7$ & $\Sigma F_\Sigma + \frac{3}{\sqrt{2}} \Delta F_\Delta - F_7$ & $\Sigma G + \frac{3}{\sqrt{2}} \Delta \tilde{b}_4^\prime - G_7$\\
    $\Sigma_c^0 \to n \gamma$  & $-\Sigma E_\Sigma^\prime + \Delta E_\Delta^\prime + E_7^\prime$ & $\sqrt{2}\Sigma F_\Sigma^\prime - \Delta F_\Delta - \sqrt{2} F_7$ & $\sqrt{2}\Sigma G^\prime - \Delta \tilde{b}_4^\prime - \sqrt{2} G_7$\\
    $\Xi_c^{\prime +} \to \Sigma^+ \gamma$  & $\Sigma E_\Sigma + \Delta E_\Delta + E_7$ & $\Sigma F_\Sigma - \frac{3}{\sqrt{2}} \Delta F_\Delta + F_7$ & $\Sigma G - \frac{3}{\sqrt{2}} \Delta \tilde{b}_4^\prime + G_7$\\
    $\Xi_c^{\prime 0} \to \Lambda \gamma$  & $\frac{1}{\sqrt{6}}\Sigma E_\Sigma^{\prime\prime} + \frac{\sqrt{3}}{2}(\Delta E_\Delta^\prime + E_7^\prime)$ & $\frac{1}{\sqrt{6}}(2F_\Sigma - F_\Sigma^\prime) - \frac{\sqrt{3}}{2}\Delta F_\Delta -\sqrt{\frac{3}{2}} F_7$ & $-\frac{1}{\sqrt{6}}(2G - G^\prime) + \frac{\sqrt{3}}{2}\Delta \tilde{b}_4^\prime +\sqrt{\frac{3}{2}} G_7$\\
    $\Xi_c^{\prime 0} \to \Sigma^0 \gamma$  & $-\frac{1}{\sqrt{2}} \Sigma E_\Sigma^{\prime\prime} + \frac{1}{2}(\Delta E_\Delta^\prime + E_7^\prime)$ & $\frac{1}{\sqrt{2}}\Sigma (2F_\Sigma - F_\Sigma^\prime) +\frac{1}{2} \Delta F_\Delta + \frac{1}{\sqrt{2}}F_7$ & $-\frac{1}{\sqrt{2}}\Sigma (2G - G^\prime) -\frac{1}{2} \Delta \tilde{b}_4^\prime - \frac{1}{\sqrt{2}}G_7$\\
    $\Omega_c \to \Xi^0 \gamma$  & $\Sigma E_\Sigma^\prime + \Delta E_\Delta^\prime + E_7^\prime$ & $\sqrt{2}\Sigma F_\Sigma^\prime + \Delta F_\Delta + \sqrt{2}F_7$ & $\sqrt{2}\Sigma G^\prime + \Delta \tilde{b}_4^\prime + \sqrt{2}G_7$\\
    \hline
    $\Xi_c^{\prime +} \to p \gamma$  & $V_{cd}^* V_{us} E_\Sigma$ & $-V_{cd}^* V_{us} F_\Sigma$ & $V_{cd}^* V_{us} G$\\
    $\Xi_c^{\prime 0} \to n \gamma$  & $V_{cd}^* V_{us} \frac{1}{\sqrt{2}}E_\Sigma^\prime$ & $-V_{cd}^* V_{us} F_\Sigma^\prime$ & $V_{cd}^* V_{us} G^\prime$\\
    $\Omega_c \to \Lambda \gamma$  & $V_{cd}^* V_{us}\left(\frac{\sqrt{3}}{2\sqrt{2}}E_\Sigma^\prime - \frac{1}{2\sqrt{3}}E_\Sigma^{\prime\prime}\right)$ & $-\sqrt{\frac{1}{3}} V_{cd}^* V_{us} (F_\Sigma + F_\Sigma^\prime)$ & $-\sqrt{\frac{1}{3}} V_{cd}^* V_{us} (G + G^\prime)$\\
    $\Omega_c \to \Sigma^0 \gamma$  & $V_{cd}^* V_{us}\left(\frac{1}{2\sqrt{2}}E_\Sigma^\prime + \frac{1}{2}E_\Sigma^{\prime\prime}\right)$ & $-V_{cd}^* V_{us} (F_\Sigma - F_\Sigma^\prime)$ & $-V_{cd}^* V_{us} (G - G^\prime)$\\
  \end{tabular}}
  \caption{Flavor symmetry relations of the decay amplitudes for the charmed sextet baryons. Analogously to Table \ref{tbl:Flavor Relationen}, $E_\Sigma^{(\prime, \prime \prime)}$, $E_\Delta^{(\prime)}$ and $E_7^{(\prime)}$ denote the U-spin triplet, U-spin singlet and the $c \to u \gamma$ contributions, respectively. Note that $\sqrt{2} E_7 = E_7^\prime$ in the isospin limit. An analogous notation is used for the $SU(3)_F$ relations with $F_\Sigma = \sqrt{\frac{2}{5}}A^\prime_{\overline{6}}- \frac{2}{3\sqrt{5}}A^\prime_{15}$, $F^\prime_\Sigma = \sqrt{\frac{2}{5}}A^\prime_{\overline{6}}+ \frac{2}{3\sqrt{5}}A^\prime_{15}$, $F_\Delta = \frac{\sqrt{2}}{3\sqrt{5}}A^\prime_{15}$ and $F_7 = \sqrt{\frac{1}{3}} (F_\text{NP} + \Delta A^\prime_3)$. Similarly, $G = -\sqrt{2} (\tilde{b}^\prime_1 - \tilde{b}^\prime_3 + \tilde{b}_4^\prime )$, $G^\prime = -\sqrt{2} (\tilde{b}^\prime_1 - \tilde{b}^\prime_3 - \tilde{b}_4^\prime) $ and $G_7 = \frac{1}{\sqrt{2}}b_1^\prime$ denote the weak annihilation and $c \to u \gamma$ contributions in the $SU(3)_F$ IRA. The $b_i, \tilde b_i$ are defined in App.~\ref{app:SU3}. The $SU(3)_F$ and U-spin decompositions are identical up to global signs for $E_\Sigma^{\prime\prime}= \frac{1}{\sqrt{2}}E_\Sigma^\prime -2 E_\Sigma$ and $E_\Delta =-\frac{3}{\sqrt{2}}E_\Delta^\prime$.}
  \label{tbl:Flavor Relationen sextet}
\end{table}
Based on the U-spin relations in Table \ref{tbl:Flavor Relationen} and \ref{tbl:Flavor Relationen sextet}, we obtain the following sum rules for the CF and SCS decays
\begin{align}
  \begin{split}
    &\A(\Lambda_c \to p \gamma) - \A(\Xi_c^+ \to \Sigma^+ \gamma) + 2\frac{\Sigma}{V_{cs}^* V_{ud}} \A(\Lambda_c \to \Sigma^+ \gamma) = 0\, , \\
    &\sqrt{3}\A(\Xi_c^0 \to \Lambda \gamma) + \A(\Xi_c^0 \to \Sigma^0 \gamma) + 2\sqrt{2}\frac{\Sigma}{V_{cs}^* V_{ud}} \A(\Xi_c^0 \to \Xi^0 \gamma) = 0\, , \\
    &\A(\Sigma_c^+ \to p \gamma) - \A(\Xi_c^{\prime +} \to \Sigma^+ \gamma) + 2\frac{\Sigma}{V_{cs}^* V_{ud}} \A(\Sigma_c^+ \to \Sigma^+ \gamma) = 0\, , \\
    &\frac{\sqrt{3}}{2}\A(\Sigma_c^0 \to n \gamma) - \A(\Xi_c^{\prime 0} \to \Lambda \gamma) + \sqrt{2}\frac{\Sigma}{V_{cs}^* V_{ud}} \A(\Sigma_c^0 \to \Lambda \gamma) = 0\, , \\
    &2\A(\Sigma_c^0 \to n \gamma) - \A(\Xi_c^{\prime 0} \to \Sigma^0 \gamma) -2\sqrt{2} \frac{\Sigma}{V_{cs}^* V_{ud}} \A(\Sigma_c^0 \to \Sigma^0 \gamma) = 0\, , \\
    &\A(\Sigma_c^0 \to n \gamma) - \A(\Omega_c \to \Xi^0 \gamma) + 2\sqrt{2}\frac{\Sigma}{V_{cs}^* V_{ud}} \A(\Xi_c^{\prime 0} \to \Xi^0 \gamma) = 0\, .
  \end{split} 
\end{align}

Flavor symmetries imply relations between the hadronic transition form factors. Using U-spin/isospin we obtain from Table \ref{tbl:Flavor Relationen}
\begin{align} \label{eq:ff3to8}
  -\sqrt{6}h_\perp^{\Xi_c^0 \to \Lambda}=\sqrt{2}h_\perp^{\Xi_c^0 \to \Sigma^0} = h_\perp^{\Xi_c^+ \to \Sigma^+} = h_\perp^{\Lambda_c \to p} \, , 
\end{align}
and  for the sextet decays from Table \ref{tbl:Flavor Relationen sextet}
\begin{align}
  h_\perp^{\Sigma_c^+ \to p} = h_\perp^{\Xi_c^{\prime +} \to \Sigma^+} = \frac{1}{\sqrt{2}} h_\perp^{\Sigma_c^0 \to n} = \sqrt{\frac{2}{3}} h_\perp^{\Xi_c^{\prime 0} \to \Lambda} = \sqrt{2} h_\perp^{\Xi_c^{\prime 0} \to \Sigma^0} = \frac{1}{\sqrt{2}} h_\perp^{\Omega_c \to \Xi^0}\, .
\end{align}
Note that other form factors based on operators with the same flavor structure, such as the form factors for semileptonic $c \to u \ell \ell$ transitions, obey the same relations.

In the computation of the $SU(3)_F$ IRA relations, detailed in App.~\ref{app:SU3}, we followed the notation of \cite{Wang:2020wxn}.
We find multiple disagreements with Table 4 of \cite{Wang:2020wxn}. However, the relations that we have determined with three different methods are consistent with each other.
Furthermore, we checked that our computational method is consistent with results for $b$-baryons \cite{Dery:2020lbc}.

Amplitudes of SCS decays can be written as $x_\Sigma \Sigma X_\Sigma +x_\Delta \Delta X_\Delta + x_7 X_7$, see Tables \ref{tbl:Flavor Relationen} and \ref{tbl:Flavor Relationen sextet}. The (relative) NP sensitivity is therefore related to $|x_7/x_\Sigma|$, which is $1,1,1/3, 1$ for the anti-triplet decays
$\Lambda_c \to p \gamma$, $\Xi_c^+ \to \Sigma^+ \gamma$, $\Xi_c^0 \to \Lambda \gamma$  and $\Xi_c^0 \to \Sigma^0 \gamma$, respectively.
$|x_7/x_\Sigma|$ is unity for all SCS sextet to octet decays except for $\Xi_c^{\prime 0} \to \Lambda \gamma$ , where it is 
$3$. The sensitivity hierarchy is therefore inverted between charmed anti-triplet and sextet baryon decays. To summarize, the hierarchies for the NP sensitivity are as follows
\begin{align}
  \begin{split}
    &\left|\frac{\A^{\text{NP}}(\Lambda_c \to p \gamma)}{\A^{\text{SM}}(\Lambda_c \to p \gamma)}\right| \approx \left|\frac{\A^{\text{NP}}(\Xi_c^+ \to \Sigma^+ \gamma)}{\A^{\text{SM}}(\Xi_c^+ \to \Sigma^+ \gamma)}\right| \approx \left|\frac{\A^{\text{NP}}(\Xi_c^0 \to \Sigma^0 \gamma)}{\A^{\text{SM}}(\Xi_c^0 \to \Sigma^0 \gamma)}\right| > \left|\frac{\A^{\text{NP}}(\Xi_c^0 \to \Lambda \gamma)}{\A^{\text{SM}}(\Xi_c^0 \to \Lambda \gamma)}\right|\, , \\ \label{eq:hierarchie}
    &\left|\frac{\A^{\text{NP}}(\Xi_c^{\prime 0} \to \Lambda \gamma)}{\A^{\text{SM}}(\Xi_c^{\prime 0} \to \Lambda \gamma)}\right| > \left|\frac{\A^{\text{NP}}(\Sigma_c^+ \to p \gamma)}{\A^{\text{SM}}(\Sigma_c^+ \to p \gamma)}\right| \approx \text{remaining sextet decay modes}\,.
  \end{split}
\end{align}

\section{Observables \label{sec:obs}}

Radiative decays of charm baryons offer interesting observables besides their branching ratios, the photon polarization and CP-asymmetry. In Sec.~\ref{sec:two body decay} we discuss the two-body decays $B_c \rightarrow B \gamma$ and how to determine the photon polarization from initially polarized charm baryons \cite{deBoer:2017que}. In Sec.~\ref{sec:decay chain}, we consider the decay chain $B_c \to B (\to B^\prime P) \gamma$ with a pseudo-scalar $P$ and present an alternative method for the experimental determination of the photon polarization
based on self-analyzing secondary decays. Relations between branching ratios of decays within the same multiplets are worked out in Sec.~\ref{sec:brs}.
In Sec.~\ref{sec:CP} we discuss the CP-asymmetries in the decay rates of SCS decays.
Possible effects of new physics are estimated in Sec.~\ref{sec:BSM reach}.

\subsection{The two-body decay $B_c \rightarrow B \gamma$}
\label{sec:two body decay}

The $B_c(P, s_{B_c}) \to B(q, s_B) \gamma(k, \epsilon^*)$ decay amplitude  (\ref{eq:amp}) can be written in terms of helicity amplitudes
\begin{align}
  \A(B_c \to B \gamma) = \frac{G_F e}{\sqrt{2}} H_1^{h_\gamma}(s_{B_c}, s_B)\, , \label{eq:amp_Bc_to_B_gamma}
\end{align}
where 
$h_\gamma$ denotes the helicity of the photon. With the explicit spinor representations from \cite{Boer:2014kda, Haber:1994pe}, the non-zero helicity amplitudes are given by
\begin{align}
  \begin{split}
    &H_1^{-1}\left(+1/2, -1/2\right) = -\sqrt{2} F_L (m_{B_c}^2 - m_B^2) \sin\left(\theta_\gamma/2\right)\, , \\
    &H_1^{-1}\left(-1/2, -1/2\right) = +\sqrt{2} F_L (m_{B_c}^2 - m_B^2) \cos\left(\theta_\gamma/2\right)\, , \\
    &H_1^{+1}\left(+1/2, +1/2\right) = +\sqrt{2} F_R (m_{B_c}^2 - m_B^2) \cos\left(\theta_\gamma/2\right)\, , \\
    &H_1^{+1}\left(-1/2, +1/2\right) = +\sqrt{2} F_R (m_{B_c}^2 - m_B^2) \sin\left(\theta_\gamma/2\right)\, , \\
  \end{split}
\end{align}
where $F_{L/R}$  denote the contributions for left-/right-handed photons (\ref{eq:amp}). $\theta_\gamma$ is the angle between the photon momentum and the quantisation axis of the spin in the $B_c$ rest frame. The decay probability is given by \cite{Legger:2006cq}
\begin{align}
  w = \frac{G_F^2 e^2}{2} \sum_{h_\gamma, s_{B_c}, s_B} \rho_{s_{B_c}, s_{B_c}} \left|H_1^{h_\gamma}(s_{B_c}, s_B)\right|^2\, , 
\end{align}
where the $B_c$ polarization is taken into account by the polarization density matrix $\rho$. The diagonal elements of $\rho$ satisfy $\rho_{+1/2,+1/2} + \rho_{-1/2,-1/2} = 1$ and define the $B_c$ polarization $P_{B_c} = \rho_{+1/2,+1/2} - \rho_{-1/2,-1/2}$. The differential branching ratio is given by
\begin{align}
  \frac{\mathrm{d}\mathcal{B}}{\mathrm{d}\cos(\theta_\gamma)} = \frac{G_F^2 e^2}{64 \pi \Gamma_{B_c}} m_{B_c}^3 \left(1 - \frac{m_B^2}{m_{B_c}^2}\right)^3 \left(|F_L|^2 + |F_R|^2\right) \left[1 + P_{B_c} \lambda_\gamma \cos(\theta_\gamma)\right]\, , \label{eq:angular dist 1}
\end{align}
where the photon polarization parameter is defined as 
\begin{align} \label{eq:lg}
  \lambda_\gamma =\frac{|F_R|^2-|F_L|^2}{|F_R|^2+|F_L|^2}= - \frac{1-r^2}{1+r^2}\,, \qquad r = \left| \frac{F_R}{F_L} \right|\, .
\end{align}
$\lambda_\gamma=-1$ corresponds to purely left-handed photons. Branching ratios  are obtained as
\begin{align}
\mathcal{B}(B_c \to B \gamma)=  \int_{-1}^{+1} \frac{\mathrm{d}\mathcal{B}}{\mathrm{d}\cos(\theta_\gamma)} \mathrm{d}\cos(\theta_\gamma)= \frac{G_F^2 e^2}{32 \pi \Gamma_{B_c}} m_{B_c}^3 \left(1 - \frac{m_B^2}{m_{B_c}^2}\right)^3 \left(|F_L|^2 + |F_R|^2\right) \, , \label{eq:br}
\end{align}
and
are not sensitive to the photon polarization parameter. However, the angular dependence allows to define a forward-backward asymmetry, which only depends on the $B_c$ polarization and $\lambda_\gamma$
\begin{align}
  A_{\text{FB}}^\gamma = \frac{1}{\mathcal{B}} \left(\int_0^1 \mathrm{d}\cos(\theta_\gamma)\frac{\mathrm{d}\mathcal{B}}{\mathrm{d}\cos(\theta_\gamma)} - \int_{-1}^0 \mathrm{d}\cos(\theta_\gamma)\frac{\mathrm{d}\mathcal{B}}{\mathrm{d}\cos(\theta_\gamma)}\right) = \frac{P_{B_c} \lambda_\gamma}{2}\, . \label{eq: AFB 1}
\end{align}
Above, $m_{B_c}$ and  $\Gamma_{B_c}$ denotes the mass and total width of the charm baryon, respectively, and $m_B$ is the mass of the secondary baryon.

$A_{\text{FB}}^\gamma$ can be related to the average longitudinal momentum $\braket{k_\parallel}_\beta$ of the photon in the lab frame with respect to the $B_c$ boost axis \cite{Hiller:2001zj}
\begin{align}
  \braket{k_\parallel}_\beta = \gamma E_\gamma (\beta + \frac{2}{3} A_{\text{FB}}^\gamma)\, ,
\end{align}
where $E_\gamma=(m_{B_c}^2 - m_B^2)/(2m_{B_c})$ is the photon energy in the $B_c$ rest frame and $\beta = |\vec{P}|/E_{B_c}$. However, a determination of the photon polarization with two-body decays is only possible if the charm baryons are polarized.

\subsection{The decay chain $B_c \to B (\to B^\prime P) \gamma$}
\label{sec:decay chain}

The $B(q, s_B) \to B^\prime(q_1, s_{B^\prime}) P(q_2)$ decay amplitude is given by \cite{Boer:2014kda}
\begin{align}
  \A(B \to B^\prime P) = N \ubar(q, s_B) \left(\xi \gamma_5 + \omega\right) u(q_1, s_{B^\prime}) = N H_2(s_B, s_{B^\prime})\, ,
\end{align}
where $P$ denotes a pseudo-scalar meson such as a pion, and $N = \frac{4G_F}{\sqrt{2}} V_{ud}^* V_{us}$ for weak hyperon decays. $\xi$ and $\omega$ are couplings of opposite parity. The helicity amplitudes can be written as \cite{Boer:2014kda}
\begin{align}
  \begin{split}
    &H_2 (+1/2,+1/2) = \left(\sqrt{r_+} \omega - \sqrt{r_-}\xi\right)\cos(\theta_B/2)\,,\\
    &H_2 (+1/2,-1/2) = \left(\sqrt{r_+} \omega + \sqrt{r_-}\xi\right)\sin(\theta_B/2)e^{i\phi_B}\,,\\
    &H_2 (-1/2,+1/2) = \left(-\sqrt{r_+} \omega + \sqrt{r_-}\xi\right)\sin(\theta_B/2)e^{-i\phi_B}\,,\\
    &H_2 (-1/2,-1/2) = \left(\sqrt{r_+} \omega + \sqrt{r_-}\xi\right)\cos(\theta_B/2)\,.
  \end{split}
\end{align}
$\theta_B$ is the angle between the $B$ flight direction and the $B^\prime$ momentum in the $B^\prime P$ rest frame. The differential branching ratio can be written as
\begin{align}
  \frac{\mathrm{d}\mathcal{B}}{\mathrm{d}\cos(\theta_B)} = \frac{|N|^2 \sqrt{r_+ r_-}}{32\pi m_B^3 \Gamma_B} \left(r_+ |\omega|^2 + r_- |\xi|^2\right)\left(1 + P_B \alpha_B \cos(\theta_B)\right)\, , \label{eq:angular dist 2}
\end{align}
with the $B$ polarization $P_B$ and the parity violating parameter
\begin{align}
  \alpha_B = \frac{-2Re(\omega^* \xi)}{\sqrt{\frac{r_-}{r_+}}|\xi|^2 + \sqrt{\frac{r_+}{r_-}}|\omega|^2}\
\end{align}
and $r_\pm = (m_B \pm m_{B^\prime})^2 - m_P^2$. For the double differential branching ratio of the decay chain $B_c \to B (\to B^\prime P) \gamma$, we obtain
\begin{align}
  \begin{split}
    \frac{\mathrm{d}^2\mathcal{B}}{\mathrm{d}\cos(\theta_\gamma) \mathrm{d}\cos(\theta_B)} \propto \left[1 + P_{B_c} \alpha_B \cos(\theta_\gamma) \cos(\theta_B) + \alpha_B \lambda_\gamma \cos(\theta_B) + P_{B_c} \lambda_\gamma \cos(\theta_\gamma)\right]\, .
  \end{split}
\end{align}
By integrating over $\theta_B$ we recover the angular dependence as in  \eqref{eq:angular dist 1}. Thus, for  polarized charm baryons one can extract $\lambda_\gamma$ via 
$A_\text{FB}^\gamma$ \eqref{eq: AFB 1}. By integrating over $\theta_\gamma$ we obtain the angular dependence as in \eqref{eq:angular dist 2}. However, the polarization of the baryon $B$ coincides with the photon polarization. Thus, as $P_B = \lambda_\gamma$, the resulting angular distribution contains a dependence on the polarization parameter, even for unpolarized $B_c$ baryons. $\lambda_\gamma$ can be determined by the  forward-backward asymmetry in the angle $\theta_B$,
\begin{equation}
  A_{\text{FB}}^B = \frac{1}{\mathcal{B}} \left(\int_0^1 \mathrm{d}\cos(\theta_B)\frac{\mathrm{d}\mathcal{B}}{\mathrm{d}\cos(\theta_B)} - \int_{-1}^0 \mathrm{d}\cos(\theta_B)\frac{\mathrm{d}\mathcal{B}}{\mathrm{d}\cos(\theta_B)}\right) = \frac{\alpha_B \lambda_\gamma}{2}\, .
\end{equation}
Decay chains with higher resonances for the secondary baryon $B$, such as $\Lambda^* \to p K$, discussed for instance in $b$-baryon decays \cite{Legger:2006cq},
can be used to study branching ratios, $A_\text{CP}$ and $A_\text{FB}^\gamma$, but not for $A_{\text{FB}}^B$ because they decay
 via the strong interaction, and have $\alpha_B$=0.

\subsection{Relating branching fractions \label{sec:brs}}

Currently, there are no experimental data on the branching fractions (\ref{eq:br})  on any of the  $B_c \to B \gamma$ decays. Using life times and phase space factors, with input compiled in App.~\ref{app:parameter},  together with
flavor symmetry one can however relate  branching ratios of decays within the same multiplet, and identify possible hierarchies between them. In the following we
assume that  branching ratios are dominated by the SM contribution, corresponding to the U-spin triplet operators (\ref{eq:Leff_SU(2)_Operatoren}).

Using Table \ref{tbl:Flavor Relationen} the  branching fractions  of the SCS decays of charmed anti-triplet baryons can be written as 
\begin{align}
  \begin{split}
    &\B(\Lambda_c \to p \gamma) \approx \lambda^2 \frac{(m_{\Lambda_c}^2 - m_p^2)^3}{(m_{\Lambda_c}^2 - m_{\Sigma^+}^2)^3} \B(\Lambda_c \to \Sigma^+ \gamma) \approx 0.072 \cdot \B(\Lambda_c \to \Sigma^+ \gamma)\, ,\\
    &\B(\Xi_c^+ \to \Sigma^+ \gamma) \approx \lambda^2 \frac{m_{\Lambda_c}^3 \Gamma_{\Lambda_c}}{m_{\Xi_c^+}^3 \Gamma_{\Xi_c^+}} \frac{(m_{\Xi^+_c}^2 - m_{\Sigma^+}^2)^3}{(m_{\Lambda_c}^2 - m_{\Sigma^+}^2)^3} \B(\Lambda_c \to \Sigma^+ \gamma) \approx 0.160 \cdot \B(\Lambda_c \to \Sigma^+ \gamma)\, ,\\
    &\B(\Xi_c^0 \to \Lambda \gamma) \approx \frac{3\lambda^2}{2} \frac{(m_{\Xi_c^0}^2 - m_\Lambda^2)^3}{(m_{\Xi_c^0}^2 - m_{\Xi^0}^2)^3} \B(\Xi_c^0 \to \Xi^0 \gamma) \approx 0.104 \cdot \B(\Xi_c^0 \to \Xi^0 \gamma)\, ,\\
    &\B(\Xi_c^0 \to \Sigma^0 \gamma) \approx \frac{\lambda^2}{2} \frac{(m_{\Xi_c^0}^2 - m_{\Sigma^0}^2)^3}{(m_{\Xi_c^0}^2 - m_{\Xi^0}^2)^3} \B(\Xi_c^0 \to \Xi^0 \gamma) \approx 0.030 \cdot \B(\Xi_c^0 \to \Xi^0 \gamma)\, .
  \end{split} \label{eq:Hierarchie BR}
\end{align}
In the $SU(3)_F$ limit, the amplitudes of the CF decays differ by the sign of the Wilson coefficient suppressed amplitude $A_{15}$. Assuming $B_\Sigma \approx B_\Sigma^\prime$ and taking into account masses and decay widths, we thus expect only small differences in the branching ratios as
\begin{align}
  \frac{\B(\Lambda_c \to \Sigma^+ \gamma)}{\B(\Xi_c^0 \to \Xi^0 \gamma)} \approx \frac{m_{\Xi_c^0}^3 \Gamma_{\Xi_c^0}}{m_{\Lambda_c}^3 \Gamma_{\Lambda_c}} \frac{(m_{\Lambda_c}^2 - m_{\Sigma^+}^2)^3}{(m_{\Xi_c^0}^2 - m_{\Xi^0}^2)^3} \approx 1.1\, .
\end{align}
Thus, the hierarchy of the SCS branching ratios can be inferred from the prefactors in \eqref{eq:Hierarchie BR}. The largest branching ratios are expected for $\Xi_c^+ \to \Sigma^+ \gamma$ decays, followed by $\Xi_c^0 \to \Lambda \gamma$ and then $\Lambda_c \to p \gamma$, all roughly about  one order of magnitude lower than the CF ones. The smallest branching ratios are obtained for $\Xi_c^0 \to \Sigma^0 \gamma$ decays.

Analogously, one obtains four simple relations for the decays of charmed sextet baryons into octet baryons using  U-spin symmetry, see Table \ref{tbl:Flavor Relationen sextet}, 
\begin{align}
  \begin{split}
    &\B(\Sigma_c^+ \to p \gamma) \approx \lambda^2 \frac{(m_{\Sigma_c^+}^2 - m_p^2)^3}{(m_{\Sigma_c^+}^2 - m_{\Sigma^+}^2)^3} \B(\Sigma_c^+ \to \Sigma^+ \gamma) \approx 0.070 \cdot \B(\Sigma_c^+ \to \Sigma^+ \gamma)\,,\\
    &\B(\Xi_c^{\prime +} \to \Sigma^+ \gamma) \approx \lambda^2 \frac{m_{\Sigma_c^+}^3 \Gamma_{\Sigma_c^+}}{m_{\Xi_c^{\prime +}}^3 \Gamma_{\Xi_c^{\prime +}}} \frac{(m_{\Xi_c^{\prime +}}^2 - m_{\Sigma^+}^2)^3}{(m_{\Sigma_c^+}^2 - m_{\Sigma^+}^2)^3} \B(\Sigma_c^+ \to \Sigma^+ \gamma)\,, \\
    &\B(\Sigma_c^0 \to n \gamma) \approx 2\lambda^2 \frac{m_{\Xi_c^{\prime 0}}^3 \Gamma_{\Xi_c^{\prime 0}}}{m_{\Sigma_c^0}^3 \Gamma_{\Sigma_c^0}} \frac{(m_{\Sigma_c^0}^2 - m_{n}^2)^3}{(m_{\Xi_c^{\prime 0}}^2 - m_{\Xi^0}^2)^3} \B(\Xi_c^{\prime 0} \to \Xi^0 \gamma)\,, \\
    &\B(\Omega_c \to \Xi^0 \gamma) \approx 2\lambda^2 \frac{m_{\Xi_c^{\prime 0}}^3 \Gamma_{\Xi_c^{\prime 0}}}{m_{\Omega_c}^3 \Gamma_{\Omega_c}} \frac{(m_{\Omega_c}^2 - m_{\Xi^0}^2)^3}{(m_{\Xi_c^{\prime 0}}^2 - m_{\Xi^0}^2)^3} \B(\Xi_c^{\prime 0} \to \Xi^0 \gamma)\,. \\
  \end{split} \label{eq:Hierarchie BR sextet}
\end{align}
The decay widths of the $\Xi_c^\prime$ are presently unknown. However, the $\Sigma_c$ and $\Xi_c^\prime$ decay strongly and electromagnetically, respectively. Thus, their branching ratios for $B_{c6} \to B_8 \gamma$ are strongly suppressed due to significantly larger total decay widths than the ones of the charm baryon anti-triplet. For e.g. $\Sigma_c \to \Sigma^+ \gamma$, there is a relative suppression of $\Gamma_{\Lambda_c}/\Gamma_{\Sigma_c^+} > 7 \cdot 10^{-10}$ compared to $\Lambda_c \to \Sigma_c^+ \gamma$. Among the sextet baryons only the $\Omega_c$ decays exclusively via the weak interaction and should therefore have significantly larger branching ratios to $B_8 \gamma$.

Note, recent analysis of charged current, semileptonic charm baryon branching ratios  suggests  large breaking of $SU(3)_F$ using simple form factor models \cite{He:2021qnc}.
It would be interesting to revisit this analysis once experimental information has become more precise and information on the dilepton spectrum has become available.

\subsection{CP-Asymmetries   \label{sec:CP}}

The CP-asymmetry in the decay rate is defined as
\begin{equation}
  A_\text{CP} = \frac{|A|^2 - |\bar{A}|^2}{|A|^2 - |\bar{A}|^2} \, .
\end{equation}
Here, $\bar A$ denotes the amplitude of the CP-conjugated decay.
In the SM, CP violation stems from the CKM matrix elements, which,  in $|\Delta c|=|\Delta u|=1$ transitions, is subjected to strong parametric suppression
\begin{equation}
  A_\text{CP}^\text{SM} \approx \text{Im}\left(\frac{-2 \Delta}{\Sigma}\right)\text{Im}\left(\frac{A_\Delta}{A_\Sigma}\right) \approx  -6 \times 10^{-4} \, \,  \text{Im}\left(\frac{A_\Delta}{A_\Sigma}\right) \, .
\end{equation}
Using Tables \ref{tbl:Flavor Relationen} and \ref{tbl:Flavor Relationen sextet}, we derive the following sum rules for the anti-triplet baryons
\begin{align}
  \begin{split}
    &A_\text{CP}(\Lambda_c \to p \gamma) + A_\text{CP}(\Xi_c^+ \to \Sigma^+ \gamma) = 0\, , \\
    &A_\text{CP}(\Xi_c^0 \to \Sigma^0 \gamma) + 3 A_\text{CP}(\Xi_c^0 \to \Lambda \gamma) = 0\, ,
  \end{split}
\end{align}
and the sextet baryons 
\begin{align}
  \begin{split}
    &A_\text{CP}(\Sigma_c^+ \to p \gamma) + A_\text{CP}(\Xi_c^{\prime+} \to \Sigma^+ \gamma) = 0\, , \\
    &A_\text{CP}(\Sigma_c^0 \to n \gamma) + A_\text{CP}(\Omega_c \to \Xi^0 \gamma) = 0\, , \\
    &A_\text{CP}(\Xi_c^{\prime 0} \to \Lambda \gamma) + 3 A_\text{CP}(\Xi_c^{\prime 0} \to \Sigma^0 \gamma) = 0\, ,
  \end{split}
\end{align}
which are valid in both $SU(2)_U$ and $SU(3)_F$.
See also \cite{Grossman:2018ptn} for a recent sum rule application to hadronic charm baryon decays.
We recall  that in $\Xi_c^0  \to \Sigma^0 \gamma$ decays the NP sensitivity is larger by a factor 3  than in $\Xi_c^0  \to \Lambda \gamma$, while it is the opposite hierarchy
in $\Xi_c^{\prime 0} \to (\Lambda,\Sigma^0) \gamma$ decays.
Beyond the SM, the CP-asymmetries obey the hierarchy of the amplitudes \eqref{eq:hierarchie}.
Furthermore, the BSM CP-asymmetries can become significantly larger. New physics in the electromagnetic dipole operators can lead to significant weak phases of \cite{Adolph:2020ema}
\begin{equation}
  |\text{Im}(C_7^{(\prime)})| \lesssim 2 \cdot 10^{-3} \, , 
\end{equation}
which include the constraints from  $\Delta A_\text{CP}$ \cite{Isidori:2012yx, LHCb:2019hro}. Neglecting the SM singlet contribution, the factor containing the weak phases becomes
\begin{equation}
  \left|\text{Im}\left(\frac{-2 C_7^{(\prime)}}{\Sigma}\right)\right| \approx 2 \times 10^{-2}\, , 
\end{equation}
which corresponds to an enhancement of CP violation by a factor of $\sim 30$ relative to the SM.
Note, in BSM models where $C_7^{(\prime)}$ is significantly larger than the coefficients of the chromomagnetic dipole operators effects in $c \to u \gamma$ can even be larger.

\section{Estimates of the BSM reach}
\label{sec:BSM reach}

In this section we work out the BSM reach in rare radiative charm baryon $B_c \to B \gamma$ modes.
As currently none of the  branching ratios nor polarization parameters of CF decays are measured we use the  benchmarks 
$\B^{\text{CF}} = 5\cdot 10^{-4} $  and $ \lambda_\gamma^{\text{CF}} = -0.5$
as input to our analysis.
Theory predictions \cite{Uppal:1993, Cheng:1994kp, Kamal:1983} for the branching ratios and polarizations are summarized in Table 25 of \cite{Cheng:2021qpd}. The branching ratios of the CF decays $\Lambda_c \to \Sigma^+ \gamma$ and $\Xi_c^0 \to \Xi^0 \gamma$ vary between $0.3 \cdot 10^{-4}$ and $3\cdot 10^{-4}$ for the different models. Note that some works did not take into account the Wilson coefficients which enhance the amplitude by $C_-$, see (\ref{eq:c4f}). The results \cite{Uppal:1993, Cheng:1994kp, Kamal:1983} for the photon polarization vary between 0.49 and -0.86. Due to these large differences, we choose to base the benchmarks for the BSM range estimates on the measured branching ratio $\B(D^0 \to K^* \gamma) = (4.1 \pm 0.7) \cdot 10^{-4}$ \cite{Zyla:2020zbs}. Note that the leading order weak annihilation amplitude for $B_c \to B \gamma$ is enhanced by $C_-/\tilde{C}$ compared to $D^0 \to V \gamma$ due to the different color structure. However, the hadronic matrix element may experience a suppression.

Once the branching ratio $\B^\text{CF}$ and photon polarization $\lambda_\gamma^\text{CF}$ for the SM-like CF decays are given, the corresponding left-handed and right-handed contributions to the decay amplitude $F_{L/R}^\text{CF}$ can be determined by
\begin{align}
  |F_L^\text{CF}| = \sqrt{\frac{\B^\text{CF}}{C^\text{CF}(1+(r^\text{CF})^2)}}\, , \qquad |F_R^\text{CF}| = r^\text{CF} \sqrt{\frac{\B^\text{CF}}{C^\text{CF}(1+(r^\text{CF})^2)}}\, ,
\end{align}
where the factor $C^\text{CF}$ is defined by 
\begin{align}
  \B^\text{CF} = C^\text{CF} (|F_L^\text{CF}|^2 + |F_R^\text{CF}|^2)
\end{align}
and can be read off  from \eqref{eq:br}. The ratio $r^\text{CF}$ is obtained from the photon polarization (\ref{eq:lg})
\begin{align}
  r^\text{CF} = \sqrt{\frac{1+\lambda_\gamma^\text{CF}}{1-\lambda_\gamma^\text{CF}}}\, .
\end{align}
Once $|F_{L/R}^\text{CF}|$ are known, we can use the flavor symmetry  relations in Table \ref{tbl:Flavor Relationen} and \ref{tbl:Flavor Relationen sextet} to determine the weak annihilation contributions for the SCS decay modes. For example, to obtain the SCS $\Xi_c^0 \to \Lambda \gamma$ WA amplitudes, one has to divide the CF  $\Xi_c^0 \to \Xi \gamma$ amplitudes by $V_{cs}^* V_{ud}$ and then multiply by $- \sqrt{\frac{3}{2}} \Sigma$. 
The signs of the SM amplitudes $F_{L/R}^\text{SCS}$ cannot be determined from the data. However, since we vary the coefficients $C_7^{(\prime)}$ from $-0.3$ to $0.3$ to estimate the BSM reach, 
see Eq.~(\ref{eq:c7range}), this does not affect our results. Alternatively, one could  also use the DCS modes to extract the weak annihilation amplitude
in an analogous manner.

For the BSM sensitive SCS decay modes the branching ratio can be written as
\begin{equation} \label{eq:Br-SCS}
  \B^\text{SCS} = C^\text{SCS} (|F_L^\text{SCS} + F_L^\text{NP}|^2 + |F_R^\text{SCS} + F_R^\text{NP}|^2)\, .
\end{equation}
The ratio of right- and left-handed amplitudes is obtained as
\begin{equation}
  r^\text{SCS} = \left| \frac{F_R^\text{SCS} + F_R^\text{NP}}{F_L^\text{SCS} + F_L^\text{NP}} \right|\,,
\end{equation}
and gives the photon polarization  (\ref{eq:lg}) in the SCS modes 
\begin{align} \lambda_\gamma^\text{SCS} =-\frac{1-(r^\text{SCS})^2}{1+(r^\text{SCS})^2} \, . 
\end{align}
The comparison of $r^\text{SCS} $ to $r^\text{CF} $ , or  $ \lambda_\gamma^\text{SCS} $ to $ \lambda_\gamma^\text{CF} $ probes  NP.

\begin{figure}[t]
  \centering
  \includegraphics[width=0.8\linewidth]{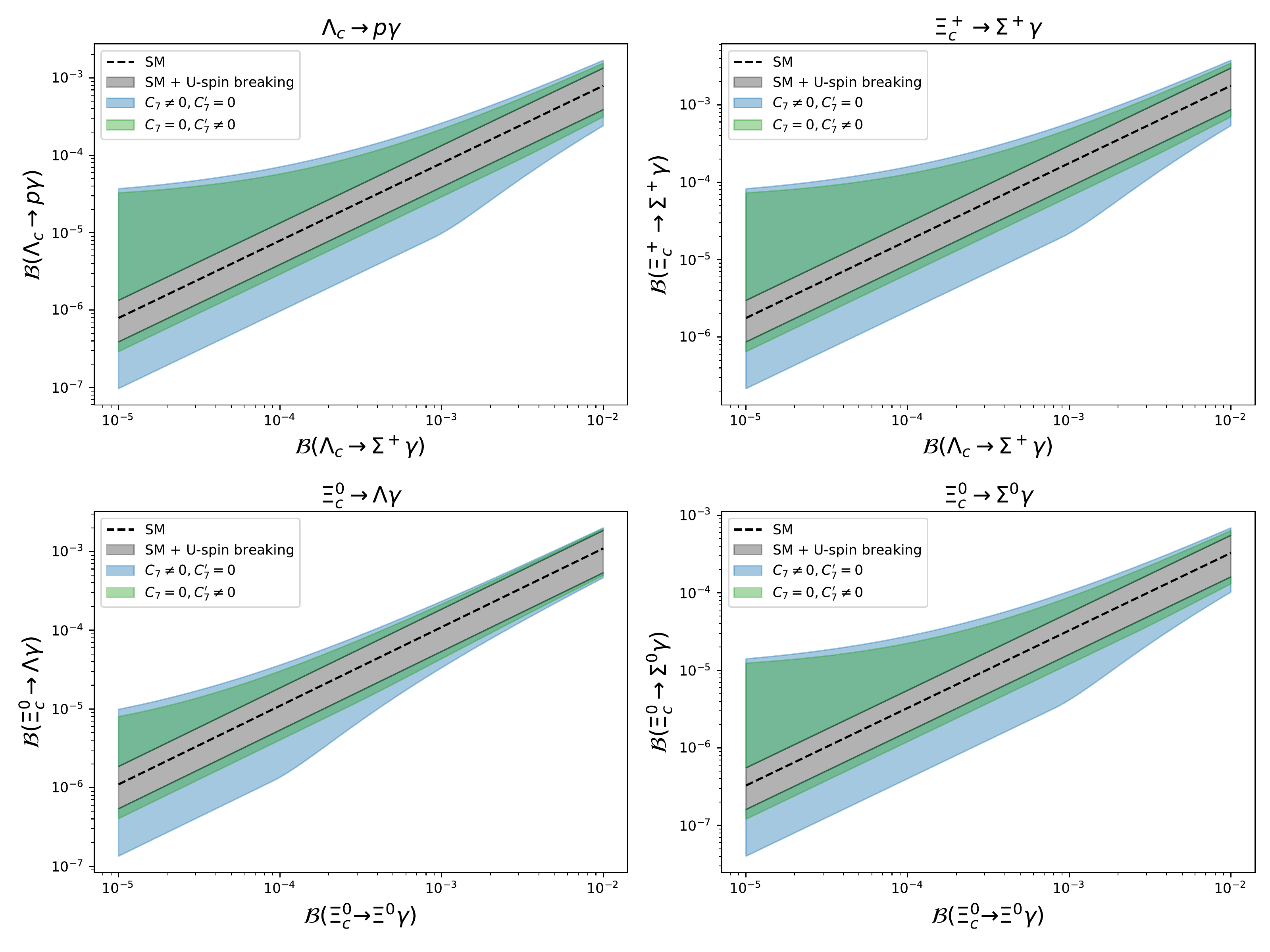}
  \caption{NP effects in the branching ratios of the BSM sensitive decay modes as a function of the branching ratios of the SM-like decay modes, for 
  $\lambda_\gamma^{\text{CF}} = -0.5$.
   The black dashed line denotes the SM in the U-spin limit. The gray shaded area shows $\pm 30\%$ U-spin breaking in $\A_{L/R}^\text{SM}$. The blue (green) region illustrates the BSM reach in $C_7$ ($C_7^\prime$). We set $C_7^\prime=0$ ($C_7=0$) and varied the other coefficient within $-0.3 \leq C_7^{(\prime)} \leq 0.3$. The BSM regions also include the $\pm 30\%$ U-spin breaking of the SM amplitudes.}
  \label{Fig:branching ratios}
\end{figure}

To begin, we first show in
Fig.~\ref{Fig:branching ratios}  the branching ratios of the BSM sensitive decays \eqref{eq:Br-SCS} as a function of the branching ratios of the CF decay modes for charmed anti-triplet baryons. The black dashed line denotes the SM in the exact U-spin limit. The gray shaded area shows $\pm 30\%$ U-spin breaking on the amplitudes $F_{L/R}^\text{SCS}$. In blue and green the BSM reach in $C_7$ (with $C_7^\prime=0$) and in $C_7^\prime$ (with $C_7=0$) is shown. The illustration of the BSM reach also includes $\pm 30\%$ U-spin breaking on the SM contributions $F_{L/R}^\text{SCS}$.
The choice for the benchmark $\lambda_\gamma^\text{CF} = -1/2$ creates an asymmetry between NP effects from $C_7$ and $C_7^\prime$.
As expected, NP can visibly affect branching ratios of SCS decays, however, a clear-cut separation from the SM is challenging in view of hadronic uncertainties.

\begin{figure}[t]
  \centering
  \includegraphics[width=0.8\linewidth]{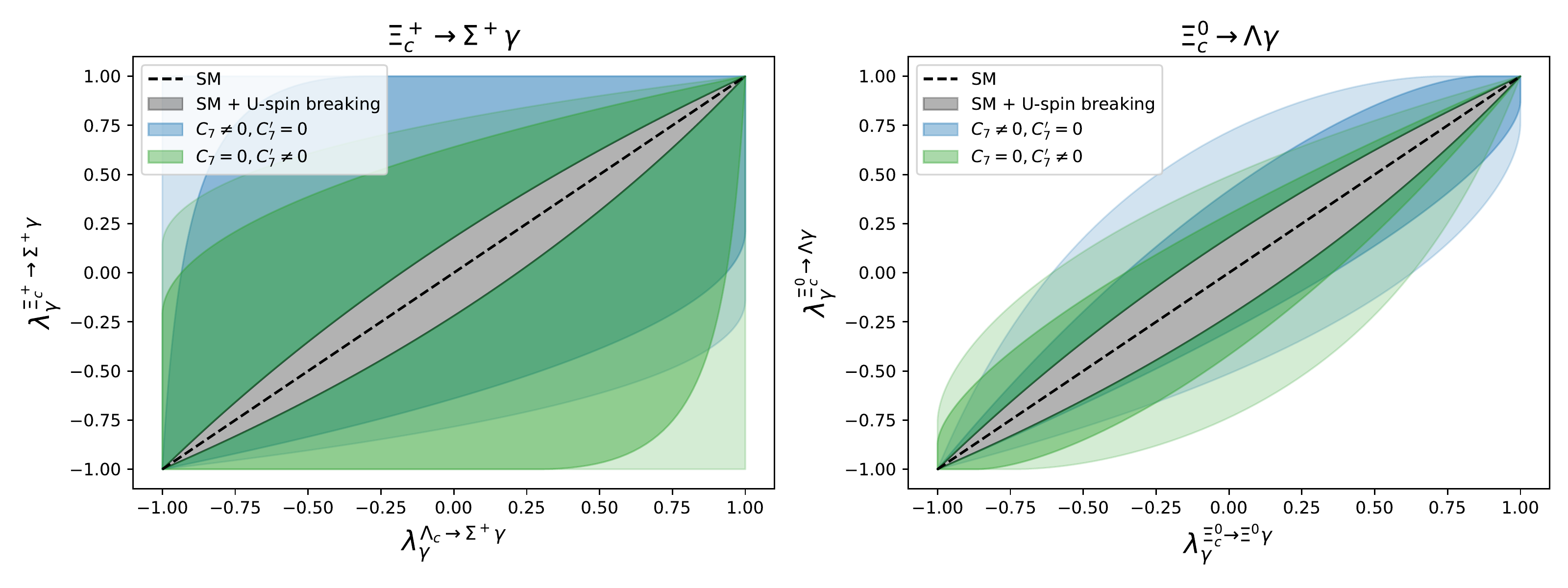}
  \caption{BSM reach of $\lambda_\gamma$ of the BSM sensitive decay modes $\Xi_c^+ \to \Sigma^+ \gamma$ (left plot) and $\Xi_c^0 \to \Lambda \gamma$ (right plot) as a function of the photon polarization of the corresponding SM-like decay modes,  $\Lambda_c \to \Sigma^+ \gamma$  and  $\Xi_c^0 \to \Xi^0 \gamma$, respectively, for  $\B^\text{CF} =5\cdot 10^{-4}$. The black dashed line denotes the SM in the  U-spin limit. The gray shaded area shows $\pm 20\%$ U-spin breaking between $r_\text{SM}^\text{CF}$ and $r_\text{SM}^\text{SCS}$. The blue (green) region illustrates the BSM reach in $C_7$ ($C_7^\prime$). We set $C_7^\prime=0$ ($C_7=0$) and varied the other coefficient within $-0.3 \leq C_7^{(\prime)} \leq 0.3$. For the darker shaded area we used the SM amplitudes in the exakt U-spin limit. For the lighter shaded area we additionally considered $\pm 30\%$ U-spin breaking in $F_{L/R}^\text{SM}$, while keeping the U-spin breaking of the ratio $r^\text{SCS}_\text{SM}$ limited to $\pm 20\%$. The BSM reach of $\Lambda_c \to p \gamma$ and $\Xi_c^0 \to \Sigma^0 \gamma$ coincides with $\Xi_c^+ \to \Sigma^+ \gamma$, see  \eqref{eq:hierarchie}, and is not shown.}
  \label{Fig:lambda_gamma}
\end{figure}

NP can be signaled in the polarization parameter, see  Fig.~\ref{Fig:lambda_gamma}. 
Here, $\lambda^{\text{SCS}}_\gamma$ is shown against $\lambda^{\text{CF}}_\gamma$.
The gray shaded area shows $\pm 20\%$ U-spin breaking between the ratios $r_\text{SM}^\text{CF}$ and $r_\text{SM}^\text{SCS}$. 
On general grounds cancellations of flavor breaking effects can be expected in ratios of amplitudes; we therefore choose a somewhat smaller range of U-spin breaking for
the ratios than for  the decay amplitudes.
The blue (green) region illustrates the BSM reach in $C_7$ ($C_7^\prime$). We set $C_7^\prime=0$ ($C_7=0$) and varied the other coefficient within $-0.3 \leq C_7^{(\prime)} \leq 0.3$.
For the darker shaded area, we varied the BSM coefficients, but used the exact U-spin limit for the SM amplitudes. 
The lighter shaded areas additionally take into account the U-spin breaking on the SM amplitudes.
As expected, the BSM reach is significantly larger for the ratio-type observable $\lambda_\gamma$ than for the branching ratio. Even with the relative suppression by a factor of 3, 
see \eqref{eq:hierarchie}, distinctive signals of new physics are possible in $\Xi_c^0 \to \Lambda \gamma$.

The sensitivity hierarchy \eqref{eq:hierarchie} can be clearly recognized in both observables, the branching ratio and the polarization parameter. Since not even the tensor form factors for the decay modes of the charmed sextet baryons are known, we do not estimate the BSM sensitivity. Note, the hierarchy of BSM sensitivity is given in \eqref{eq:hierarchie}.

Finally, in Table \ref{tbl:partner decays} we provide a list of partner decays of CF and SCS decays, which allow to probe NP by the photon polarization using the decay chains $B_c \to B (\to B^\prime P) \gamma$ and the U-spin relations given in Table \ref{tbl:Flavor Relationen} and \ref{tbl:Flavor Relationen sextet}. Two partner modes are required for 
the SCS $\Xi_c^{\prime 0} \to \Lambda \gamma$ decay to determine the SM amplitude $E_\Sigma^{\prime \prime}$.
\begin{table}[ht]
  \centering\begin{tabular}{c|c}
    BSM sensitive (SCS) decay & CF decay \\
    \hline
    $\Xi_c^+ \to \Sigma^+ \gamma$ & $\Lambda_c \to \Sigma^+ \gamma$ \\
    $\Xi_c^0 \to \Lambda \gamma$ & $\Xi_c^0 \to \Xi^0 \gamma$ \\
     $\Omega_c \to \Xi^0 \gamma$ & $\Xi_c^{\prime 0} \to \Xi^0 \gamma$\\ 
    $\Xi_c^{\prime +} \to \Sigma^+ \gamma$ & $\Sigma_c^+ \to \Sigma^+ \gamma$ \\
    $\Xi_c^{\prime 0} \to \Lambda \gamma$ & $\Sigma_c^0 \to \Lambda \gamma$, $\Xi_c^{\prime 0} \to \Xi^0 \gamma$
  \end{tabular}
  \caption{Partner modes  which enable tests of the SM by the photon polarization using  the decay chains $B_c \to B (\to B^\prime P) \gamma$ and U-spin relations given in Table \ref{tbl:Flavor Relationen} and \ref{tbl:Flavor Relationen sextet}. All secondary baryons $(\Lambda,\Sigma^+,\Xi^0)$ are self-analyzing, see Table \ref{tab:hyperons}. }
  \label{tbl:partner decays}
\end{table}
Note that the decays of $\Sigma_c$ and $\Xi_c^\prime$ are probably unusable due to their sizable decays widths, see App.~\ref{app:parameter}, and corresponding
suppression of the rare radiative decay channels. Dropping the $\Sigma_c$ and $\Xi_c^\prime$ and considering only  the decays of the $\Omega_c$ from the sextet baryons, we need both DCS decays $\Omega_c \to \Lambda \gamma$ and $\Omega_c \to \Sigma^0 \gamma$ to determine the SM amplitude of the SCS decay mode $\Omega_c \to \Xi^0 \gamma$. However, as the $\Sigma^0$ decays electromagnetically, it is not possible to determine the polarization of the photon by the decay chain $B_c \to B (\to B^\prime P) \gamma$. Thus, we need polarized $\Omega_c$ to extract the polarization of the photon. To summarize, the decays of the charmed sextet baryons are disadvantageous compared to the charmed anti-triplet baryons.

\section{Conclusions \label{sec:con}}

A key ingredient to BSM searches with radiative rare charm decays are flavor symmetries, which allow to experimentally extract
requisite SM amplitudes which are otherwise not available with sufficient accuracy.  
While the U-spin or $SU(3)_F$-ansatz are subject to systematic uncertainties up to order 30 percent, and no precision tools,
given the present situation with large room for NP in radiative $|\Delta c|=|\Delta u|=1$ transitions  they  suffice to signal NP.

Here we propose to test the SM by measuring the photon polarization in  a CF and its SCS partner mode using decay chains $B_c \to B (\to B^\prime P) \gamma$ 
with self-analyzing decay of the secondary baryon.
Comparison of the two measurements provides a data-driven null test.
In Table \ref{tbl:partner decays} we provide a list of partner decays of CF and SCS decays.

Two suitable BSM sensitive modes  can be identified 
$\Xi_c^+ \to \Sigma^+ (\to p \pi^0) \gamma$ and $\Xi_c^0 \to \Lambda (\to p \pi^-) \gamma$.
An experimental determination of polarization  asymmetries  in corresponding CF  decays  $\Lambda_c^+ \to \Sigma^+ (\to p \pi^0) \gamma$ and  $\Xi_c^0 \to \Xi^0 (\to \Lambda \pi^0) \gamma$ serves as an estimate of the ones in SCS decays.
The NP reach of these data-driven null tests is illustrated  in Fig.~\ref{Fig:lambda_gamma}. NP can be signaled in sizable deviation from the SM prediction.

The self-analyzing baryon decays offer yet another way to probe the photon polarization in FCNC transitions of charmed baryons in addition to the methods proposed in \cite{deBoer:2017que,deBoer:2018zhz,Adolph:2018hde}.
Additional SCS modes can be used as NP probes  if the initial cham baryons are polarized, including $\Lambda_c \to p \gamma$ together with  $\Lambda_c \to \Sigma^+ \gamma$ (CF) or
$\Xi_c^+ \to p \gamma$ (DCS), and $\Omega_c \to \Xi^0 \gamma$ (SCS) together with both DCS modes $\Omega_c \to \Lambda \gamma$ and $\Omega_c \to \Sigma^0 \gamma$.

We stress that none of the radiative charm baryon modes have been observed yet, a point which we hope  triggers experimental interest.
Corresponding branching ratios of rare radiative $D$-mesons are observed at the level few $\times 10^{-4}$ (CF) and $10^{-5}$ (SCS)  \cite{Zyla:2020zbs}.

\section*{Acknowledgements}
We are happy to thank  Stefan Schacht for useful discussions.
GH is grateful
to the organizers and participants of the MIAPP-program "Charming Clues for Existence"  for useful discussions.
This research was supported by the Munich Institute for Astro- and Particle Physics (MIAPP) which is funded by the Deutsche Forschungsgemeinschaft (DFG, German Research Foundation) under Germany's Excellence Strategy - EXC-2094 - 390783311.
This work is supported in part by the \textit{Bundesministerium f\"ur Bildung und Forschung} (BMBF) under project number
 05H21PECL2.

\appendix

\section{Parameters}
\label{app:parameter}
Masses and  mean life times are taken from the PDG \cite{Zyla:2020zbs}
\begin{equation*} \label{eq:data}
  \begin{alignedat}{2}
    &m_{\Lambda_c} = (2.28646 \pm 0.00014)\, \text{GeV}\, , \quad && \tau_{\Lambda_c}=(2.024 \pm 0.031) \cdot 10^{-13} \, \text{s}\, ,\\
    &m_{\Xi_c^+} = (2.46794^{+0.00017}_{-0.00020})\, \text{GeV}\, , \quad && \tau_{\Xi_c^+}=(4.56 \pm 0.05) \cdot 10^{-13} \, \text{s}\, ,\\
    &m_{\Xi_c^0} = (2.47090^{+0.00022}_{-0.00029})\, \text{GeV}\, , \quad && \tau_{\Xi_c^0}=(1.53 \pm 0.06) \cdot 10^{-13} \, \text{s}\, ,\\
    &m_{\Sigma_c^{++}} = (2.45397 \pm 0.00014)\, \text{GeV}\, , \quad && \Gamma_{\Sigma_c^{++}}=(0.00185 \pm 0.00014)  \, \GeV\, ,\\
    &m_{\Sigma_c^+} = (2.4529 \pm 0.0004)\, \text{GeV}\, , \quad && \Gamma_{\Sigma_c^+} < 0.0046  \, \GeV\, ,\\
    &m_{\Sigma_c^0} = (2.45375 \pm 0.00014)\, \text{GeV}\, , \quad && \Gamma_{\Sigma_c^0}=(0.00179 \pm 0.00015)  \, \GeV\, ,\\
    &m_{\Xi_c^{\prime +}} = (2.5782 \pm 0.0005)\, \text{GeV}\, , \quad && \, \\
    &m_{\Xi_c^{\prime 0}} = (2.5787 \pm 0.0005)\, \text{GeV}\, , \quad && \, \\
    &m_{\Omega_c} = (2.6952 \pm 0.0017)\, \text{GeV}\, , \quad && \tau_{\Omega_c}=(2.68 \pm 0.026) \cdot 10^{-13} \, \text{s}\, ,\\
    &m_\Lambda = (1.115683 \pm 0.000006) \, \text{GeV}\, , \quad &&  \tau_{\Lambda}=(2.632 \pm 0.020) \cdot 10^{-10} \, \text{s}\, ,\\
    &m_{\Sigma^+} = (1.18937 \pm 0.000007) \, \text{GeV}\, , \quad &&  \tau_{\Sigma^+}=(8.018 \pm 0.026) \cdot 10^{-11} \, \text{s}\, ,\\
    &m_p = (0.938272081 \pm 0.000000006)\, \text{GeV}\, , \quad &&  m_{\Sigma^0} = (1.192642 \pm 0.000024) \, \text{GeV}\, ,\\
    \end{alignedat}
\end{equation*}
The CKM matrix elements are taken from the UTfit collaboration \cite{UTfit}
\begin{align*}
	V_{ud} &= 0.97431 \pm 0.00012,  \quad \quad 
	V_{us} = 0.22514 \pm 0.00055, \\
	V_{cd} &= (-0.22500 \pm 0.00054) \exp \left[ \im (0.0351 \pm 0.0010)^{\circ} \right], \\
	V_{cs} &= (0.97344 \pm 0.00012) \exp \left[ \im (-0.001880 \pm 0.000055)^{\circ} \right].
\end{align*}
The decay constants are given by \cite{Straub:2015ica}
\begin{align*}
  f_\phi = (0.233 \pm 0.004)\, \GeV \, , \quad f_\omega^{(d)} = (0.2013 \pm 0.0008)\, \GeV \, , \quad f_\rho^{(d)} = (0.2097 \pm 0.0003)\, \GeV\, .
\end{align*}

\section{$SU(2)$ decomposition of decay amplitudes}
\label{app:flavor symmetrien}

In this appendix we provide additional information on the $SU(2)$ decomposition of the anti-triplet to octet and sextet to octet decay amplitudes.
Decompositions of charm sextet to decuplet baryons are given in App.~\ref{app:6to10}. In Table \ref{tbl:SU(2) states} we list the iso-, U- and V-spin wave functions of quarks and baryons. In Table \ref{tbl:U-Spin} and \ref{tbl:U-Spin sextet}, we show the U-spin decompositions of the $B_{c3} \to B_8 \gamma$ and $B_{c6} \to B_8 \gamma$ amplitudes.
For the description of the amplitudes of the U-spin triplet operators, we need two amplitudes in the case of charmed anti-triplet baryons. For the charmed sextet baryons we need 
a third amplitude. This is due to the fact that we have U-spin triplets instead of U-spin singlets in the initial state. This allows both the singlet and triplet components of the $\Lambda$ and $\Sigma^0$ to contribute.
\begin{table}[h]
  \centering
  \resizebox*{!}{0.75\textheight}{%
  \begin{tabular}{c|c|c|c|c|c}
    Particle & Quark content & $SU(3)_F$ & Isospin & U-Spin & V-spin\\
    \hline
    $\ket{u}$         & u       & $\ket{\textbf{3}, \frac{1}{3}, \frac{1}{2}, \frac{1}{2}}$ & $\ket{\frac{1}{2}, \frac{1}{2}}_I$    & $\ket{0,0}_U$                       & $\ket{\frac{1}{2}, \frac{1}{2}}_V$ \\
    $\ket{d}$         & d       & $\ket{\textbf{3}, \frac{1}{3}, \frac{1}{2}, -\frac{1}{2}}$ & $\ket{\frac{1}{2}, -\frac{1}{2}}_I$   & $\ket{\frac{1}{2}, \frac{1}{2}}_U$  & $\ket{0,0}_V$ \\
    $\ket{s}$         & s       & $\ket{\textbf{3}, -\frac{2}{3}, 0, 0}$ & $\ket{0,0}_I$                         & $\ket{\frac{1}{2}, -\frac{1}{2}}_U$ & $\ket{\frac{1}{2}, -\frac{1}{2}}_V$ \\
    $\ket{\ubar}$     & $\ubar$ & $\ket{\overline{\textbf{3}}, -\frac{1}{3}, \frac{1}{2}, -\frac{1}{2}}$ & $\ket{\frac{1}{2}, -\frac{1}{2}}_I$   & $\ket{0,0}_U$                       & $\ket{\frac{1}{2}, -\frac{1}{2}}_V$ \\
    $\ket{\dbar}$     & $\dbar$ & $\ket{\overline{\textbf{3}}, -\frac{1}{3}, \frac{1}{2}, \frac{1}{2}}$ & $-\ket{\frac{1}{2}, \frac{1}{2}}_I$   & $\ket{\frac{1}{2}, -\frac{1}{2}}_U$ & $\ket{0,0}_V$ \\
    $\ket{\sbar}$     & $\sbar$ & $\ket{\overline{\textbf{3}}, \frac{2}{3}, 0, 0}$ & $\ket{0,0}_I$                         & $-\ket{\frac{1}{2}, \frac{1}{2}}_U$ & $-\ket{\frac{1}{2}, \frac{1}{2}}_V$ \\
    \hline
    $\ket{\Lambda_c}$ & cud & $\ket{\overline{\textbf{3}}, \frac{2}{3}, 0, 0}$                       & $\ket{0,0}_I$                       & $\ket{\frac{1}{2}, \frac{1}{2}}_U$  & $\ket{\frac{1}{2}, \frac{1}{2}}_V$ \\
    $\ket{\Xi_c^0}$   & cds & $\ket{\overline{\textbf{3}}, -\frac{1}{3}, \frac{1}{2}, -\frac{1}{2}}$ & $\ket{\frac{1}{2}, -\frac{1}{2}}_I$ & $\ket{0,0}_U$                       & $\ket{\frac{1}{2}, -\frac{1}{2}}_V$ \\
    $\ket{\Xi_c^+}$   & cus & $\ket{\overline{\textbf{3}}, -\frac{1}{3}, \frac{1}{2}, +\frac{1}{2}}$ & $\ket{\frac{1}{2}, \frac{1}{2}}_I$  & $\ket{\frac{1}{2}, -\frac{1}{2}}_U$ & $\ket{0,0}_V$ \\
    \hline
    $\ket{\Omega_c}$           & css & $\ket{\textbf{6}, -\frac{4}{3}, 0, 0}$ & $\ket{0,0}_I$                       & $\ket{1,-1}_U$                      & $\ket{1,-1}_V$ \\
    $\ket{\Xi_c^{\prime 0}}$   & cds & $\ket{\textbf{6}, -\frac{1}{3}, \frac{1}{2}, -\frac{1}{2}}$ & $\ket{\frac{1}{2}, -\frac{1}{2}}_I$ & $\ket{1,0}_U$                       & $\ket{\frac{1}{2}, -\frac{1}{2}}_V$ \\
    $\ket{\Xi_c^{\prime +}}$   & cus & $\ket{\textbf{6}, -\frac{1}{3}, \frac{1}{2}, \frac{1}{2}}$ & $\ket{\frac{1}{2}, \frac{1}{2}}_I$  & $\ket{\frac{1}{2}, -\frac{1}{2}}_U$ & $\ket{1,0}_V$ \\
    $\ket{\Sigma_c^0}$         & cdd & $\ket{\textbf{6}, \frac{2}{3}, 1,-1}$ & $\ket{1,-1}_I$                      & $\ket{1,1}_U$                       & $\ket{0,0}_V$ \\
    $\ket{\Sigma_c^+}$         & cud & $\ket{\textbf{6}, \frac{2}{3}, 1,0}$ & $\ket{1,0}_I$                       & $\ket{\frac{1}{2}, \frac{1}{2}}_U$  & $\ket{\frac{1}{2}, \frac{1}{2}}_V$ \\
    $\ket{\Sigma_c^{++}}$      & cuu & $\ket{\textbf{6}, \frac{2}{3}, 1,1}$ & $\ket{1,1}_I$                       & $\ket{0,0}_U$                       & $\ket{1,1}_V$ \\
    \hline
    $\ket{\Lambda}$  & uds & $\ket{\textbf{8}, 0, 0, 0}$ & $\ket{0,0}_I$ & $\frac{\sqrt{3}}{2}\ket{1,0}_U - \frac{1}{2}\ket{0,0}_U$ & $-\frac{\sqrt{3}}{2}\ket{1,0}_V + \frac{1}{2}\ket{0,0}_V$ \\
    $\ket{\Sigma^0}$ & uds & $\ket{\textbf{8}, 0, 1, 0}$ & $\ket{1,0}_I$ & $\frac{1}{2}\ket{1,0}_U + \frac{\sqrt{3}}{2}\ket{0,0}_U$ & $\frac{1}{2}\ket{1,0}_V + \frac{\sqrt{3}}{2}\ket{0,0}_V$ \\
    $\ket{\Sigma^-}$ & dds & $\ket{\textbf{8}, 0, 1, -1}$ & $\ket{1,-1}_I$                        & $\ket{\frac{1}{2}, \frac{1}{2}}^1_U$  & $\ket{\frac{1}{2}, -\frac{1}{2}}^1_V$ \\
    $\ket{\Sigma^+}$ & uus & $\ket{\textbf{8}, 0, 1, 1}$ & $\ket{1,1}_I$                         & $\ket{\frac{1}{2}, -\frac{1}{2}}^2_U$ & $\ket{\frac{1}{2}, \frac{1}{2}}^2_V$ \\
    $\ket{\Xi^0}$    & uss & $\ket{\textbf{8}, -1, \frac{1}{2}, \frac{1}{2}}$ & $\ket{\frac{1}{2}, \frac{1}{2}}^1_I$  & $\ket{1,-1}_U$                        & $\ket{\frac{1}{2}, -\frac{1}{2}}^2_V$ \\
    $\ket{\Xi^-}$    & dss & $\ket{\textbf{8}, -1, \frac{1}{2}, -\frac{1}{2}}$ & $\ket{\frac{1}{2}, -\frac{1}{2}}^1_I$ & $\ket{\frac{1}{2}, -\frac{1}{2}}^1_U$ & $\ket{1,-1}_V$ \\
    $\ket{n}$        & udd & $\ket{\textbf{8}, 1, \frac{1}{2}, -\frac{1}{2}}$ & $\ket{\frac{1}{2}, -\frac{1}{2}}^2_I$ & $\ket{1,1}_U$                         & $\ket{\frac{1}{2}, \frac{1}{2}}^1_V$ \\
    $\ket{p}$        & uud & $\ket{\textbf{8}, 1, \frac{1}{2}, \frac{1}{2}}$ & $\ket{\frac{1}{2}, \frac{1}{2}}^2_I$  & $\ket{\frac{1}{2}, \frac{1}{2}}^2_U$  & $\ket{1,1}_V$ \\
    \hline
    $\ket{\Delta^{++}}$ & uuu & $\ket{\textbf{10}, 1, \frac{3}{2},\frac{3}{2}}$ & $\ket{\frac{3}{2},\frac{3}{2}}_I$   & $\ket{0,0}_U$                       & $\ket{\frac{3}{2},\frac{3}{2}}_V$ \\
    $\ket{\Delta^+}$    & uud & $\ket{\textbf{10}, 1, \frac{3}{2},\frac{1}{2}}$ & $\ket{\frac{3}{2},\frac{1}{2}}_I$   & $\ket{\frac{1}{2}, \frac{1}{2}}_U$  & $\ket{1,1}_V$ \\
    $\ket{\Delta^0}$    & udd & $\ket{\textbf{10}, 1, \frac{3}{2},-\frac{1}{2}}$ & $\ket{\frac{3}{2},-\frac{1}{2}}_I$  & $\ket{1,1}_U$                       & $\ket{\frac{1}{2}, \frac{1}{2}}_V$ \\
    $\ket{\Delta^-}$    & ddd & $\ket{\textbf{10}, 1, \frac{3}{2},-\frac{3}{2}}$ & $\ket{\frac{3}{2},-\frac{3}{2}}_I$  & $\ket{\frac{3}{2},\frac{3}{2}}_U$   & $\ket{0,0}_V$ \\
    $\ket{\Sigma^{*+}}$ & uus & $\ket{\textbf{10}, 0, 1, 1}$ & $\ket{1,1}_I$                       & $\ket{\frac{1}{2}, -\frac{1}{2}}_U$ & $\ket{\frac{3}{2}, \frac{1}{2}}_V$ \\
    $\ket{\Sigma^{*0}}$ & uds & $\ket{\textbf{10}, 0, 1, 0}$ & $\ket{1,0}_I$                       & $\ket{1,0}_U$                       & $\ket{1,0}_V$ \\
    $\ket{\Sigma^{*-}}$ & dds & $\ket{\textbf{10}, 0, 1, -1}$ & $\ket{1,-1}_I$                      & $\ket{\frac{3}{2},\frac{1}{2}}_U$   & $\ket{\frac{1}{2}, -\frac{1}{2}}_V$ \\
    $\ket{\Xi^{*0}}$    & uss & $\ket{\textbf{10}, -1, \frac{1}{2}, \frac{1}{2}}$ & $\ket{\frac{1}{2}, \frac{1}{2}}_I$  & $\ket{1,-1}_U$                      & $\ket{\frac{3}{2}, -\frac{1}{2}}_V$ \\
    $\ket{\Xi^{*-}}$    & dss & $\ket{\textbf{10}, -1, \frac{1}{2}, -\frac{1}{2}}$ & $\ket{\frac{1}{2}, -\frac{1}{2}}_I$ & $\ket{\frac{3}{2}, -\frac{1}{2}}_U$ & $\ket{1,-1}_V$ \\
    $\ket{\Omega^-}$    & sss & $\ket{\textbf{10}, -2, 0, 0}$ & $\ket{0,0}_I$                       & $\ket{\frac{3}{2}, -\frac{3}{2}}_U$ & $\ket{\frac{3}{2}, -\frac{3}{2}}_V$ \\
  \end{tabular}}
  \caption{Isospin, $U$-spin, $V$-spin and $SU(3)_F$ wave functions of quarks, charmed anti-triplet/sextet baryons and the light baryon octet and decuplet, analogously to \cite{Dery:2020lbc}. The superscript $1,2$ refer to different doublets within the baryon octet.}
  \label{tbl:SU(2) states}
\end{table} 
\FloatBarrier
\begin{table}[th]
  \resizebox*{!}{0.28\textheight}{%
  \centering\begin{tabular}{l|c|c|c|c}
    Decay & $\braket{\frac{1}{2}|1|\frac{1}{2}}$ & $\braket{1|1|0}$ & $\braket{\frac{1}{2}|0|\frac{1}{2}}$ & $\braket{0|0|0}$\\
    \hline
    $\Lambda_c \to \Sigma^+ \gamma$ & $\sqrt{\frac{2}{3}}V_{cs}^* V_{ud}$ & 0 & - & - \\
    $\Xi_c^0 \to \Xi^0 \gamma$ & 0 & $-V_{cs}^* V_{ud}$ & - & - \\
    \hline
    $\Lambda_c \to p \gamma$ & $-\sqrt{\frac{2}{3}}\Sigma$ & 0 & $\sqrt{2} \Delta$ & 0 \\
    $\Xi_c^+ \to \Sigma^+ \gamma$ & $\sqrt{\frac{2}{3}}\Sigma$ & 0 & $\sqrt{2} \Delta$ & 0 \\
    $\Xi_c^0 \to \Lambda \gamma$ & 0 & $\sqrt{\frac{3}{2}}\Sigma$ & 0 & $-\frac{1}{\sqrt{2}} \Delta$ \\
    $\Xi_c^0 \to \Sigma^0 \gamma$ & 0 & $\frac{1}{\sqrt{2}}\Sigma$ & 0 & $\frac{\sqrt{3}}{\sqrt{2}}\Delta$ \\
    \hline
    $\Xi_c^+ \to p \gamma$ & $\sqrt{\frac{2}{3}}V_{cd}^* V_{us}$ & 0 & - & - \\
    $\Xi_c^0 \to n \gamma$ & 0 & $V_{cd}^* V_{us}$ & - & -
  \end{tabular}}
  \caption{U-Spin decomposition of the SM decay amplitudes for the charmed anti-triplet baryons. In the matrix element $\braket{ U(f)| U(O) |  U(i)}$, $ U(f)$, $ U(O)$ and $ U(i)$ denote the $U$-spin of the final state, the U-spin changing operators and the initial state, respectively. The matrix elements are related to the amplitudes in Table \ref{tbl:Flavor Relationen} as follows: $A_\Sigma \sim \sqrt{\frac{2}{3}}\braket{\frac{1}{2}|1|\frac{1}{2}}$, $A_\Sigma^\prime \sim -\braket{1|1|0}$, $A_\Delta \sim \sqrt{2}\braket{\frac{1}{2}|0|\frac{1}{2}}$ and $A_\Delta^\prime \sim \sqrt{2}\braket{0|0|0}$.}
  \label{tbl:U-Spin}
\end{table}
\begin{table}[h!]
  \centering
  \resizebox*{!}{0.41\textheight}{%
  \begin{tabular}{l|c|c|c|c|c}
    Decay & $\braket{\frac{1}{2}|1|\frac{1}{2}}$ & $\braket{1|1|1}$ & $\braket{0|1|1}$ & $\braket{\frac{1}{2}|0|\frac{1}{2}}$ & $\braket{1|0|1}$\\
    \hline
    $\Sigma_c^+ \to \Sigma^+ \gamma$ & $\sqrt{\frac{2}{3}}V_{cs}^* V_{ud}$ & 0 & 0 & - & - \\
    $\Sigma_c^0 \to \Lambda \gamma$ & 0 & $\frac{\sqrt{3}}{2\sqrt{2}}V_{cs}^* V_{ud}$ & $\frac{1}{2\sqrt{3}}V_{cs}^* V_{ud}$ & - & - \\
    $\Sigma_c^0 \to \Sigma^0 \gamma$ & 0 & $\frac{1}{2\sqrt{2}}V_{cs}^* V_{ud}$ & $-\frac{1}{2}V_{cs}^* V_{ud}$ & - & - \\
    $\Xi_c^{\prime 0} \to \Xi^0 \gamma$ & 0 & $\frac{1}{\sqrt{2}}V_{cs}^* V_{ud}$ & 0 & - & - \\
    \hline
    $\Sigma_c^+ \to p \gamma$ & $-\sqrt{\frac{2}{3}}\Sigma$ & 0 & 0 & $\sqrt{2}\Delta$ & 0 \\
    $\Sigma_c^0 \to n \gamma$ & 0 & $-\Sigma$ & 0 & 0 & $\sqrt{2}\Delta$ \\
    $\Xi_c^{\prime +} \to \Sigma^+ \gamma$ & $\sqrt{\frac{2}{3}}\Sigma$ & 0 & 0 & $\sqrt{2}\Delta$ & 0 \\
    $\Xi_c^{\prime 0} \to \Lambda \gamma$ & 0 & 0 & $\frac{1}{\sqrt{6}}\Sigma$ & 0 & $\sqrt{\frac{3}{2}}\Delta$ \\
    $\Xi_c^{\prime 0} \to \Sigma^0 \gamma$ & 0 & 0 & $-\frac{1}{\sqrt{2}}\Sigma$ & 0 & $\frac{1}{\sqrt{2}}\Delta$ \\
    $\Omega_c \to \Xi^0 \gamma$ & 0 & $\Sigma$ & 0 & 0 & $\sqrt{2}\Delta$ \\
    \hline
    $\Xi_c^{\prime +} \to p \gamma$ & $\sqrt{\frac{2}{3}}V_{cd}^* V_{us}$ & 0 & 0 & - & - \\
    $\Xi_c^{\prime 0} \to n \gamma$ & 0 & $\frac{1}{\sqrt{2}}V_{cd}^* V_{us}$ & 0 & - & -  \\
    $\Omega_c \to \Lambda \gamma$ & 0 & $\frac{\sqrt{3}}{2\sqrt{2}}V_{cd}^* V_{us}$ & $-\frac{1}{2\sqrt{3}}V_{cd}^* V_{us}$ & - & -  \\
    $\Omega_c \to \Sigma^0 \gamma$ & 0 & $\frac{1}{2\sqrt{2}}V_{cd}^* V_{us}$ & $\frac{1}{2}V_{cd}^* V_{us}$ & - & -  
  \end{tabular}}
  \caption{U-Spin decomposition of the SM decay amplitudes for the charmed sextet baryons. In the matrix element $\braket{ U(f)| U(O) | U(i)}$, $U(f)$, $U(O)$ and $U(i)$ denote the U-spin of the final state, the U-spin changing operators and the initial state, respectively. The matrix elements are related to the amplitudes in Table \ref{tbl:Flavor Relationen sextet} as follows: $E_\Sigma \sim \sqrt{\frac{2}{3}}\braket{\frac{1}{2}|1|\frac{1}{2}}$, $E_\Sigma^\prime \sim \braket{1|1|1}$, $E_\Sigma^{\prime\prime} \sim \braket{0|1|1}$, $E_\Delta \sim \sqrt{2}\braket{\frac{1}{2}|1|\frac{1}{2}}$ and $E_\Delta^\prime \sim \sqrt{2}\braket{0|0|0}$.}
  \label{tbl:U-Spin sextet}
\end{table}
\FloatBarrier
In Table \ref{tbl:Iso- U-spin BSM} and \ref{tbl:Iso- U-spin BSM}, we show the Iso- and U-spin decomposition of the $c \to u \gamma$ contributions. Note that the $c \to u \gamma$ contributions have the same U-spin structure as the SM singlet operator. On the one hand, we distinguish between them as we study possible BSM effects in the electromagnetic dipole operators. On the other hand, differentiation enables us to use the additional simple isospin relations between the amplitudes of the dipole operators.
\begin{table}[th]
  \centering\begin{tabular}{l|c|c|c}
    Decay &$\braket{1|\frac{1}{2}|\frac{1}{2}}_I$ & $\braket{\frac{1}{2}|0|\frac{1}{2}}_U$ & $\braket{0|0|0}_U$ \\
    \hline
    $\Lambda_c \to p \gamma$ & $1$ & $1$ & $0$\\
    $\Xi_c^+ \to \Sigma^+ \gamma$ & $0$ & $1$ & $0$ \\
    $\Xi_c^0 \to \Lambda \gamma$ & $0$ & $0$ & $-\frac{1}{2}$\\
    $\Xi_c^0 \to \Sigma^0 \gamma$ & $\sqrt{\frac{1}{2}}$ & $0$ & $\frac{\sqrt{3}}{2}$
  \end{tabular}
  \caption{Iso- and U-spin decomposition of the BSM contribution for the charm anti-triplet baryons. The matrix elements are related to the amplitudes in Table \ref{tbl:Flavor Relationen} as follows: $A_7^\prime \sim \braket{\frac{1}{2}|0|\frac{1}{2}}_U$ and $A_7 \sim \braket{0|0|0}_U$}
  \label{tbl:Iso- U-spin BSM}
\end{table}
\begin{table}[th]
  \centering\begin{tabular}{l|c|c|c}
    Decay &$\braket{\frac{1}{2}|\frac{1}{2}|1}_I$ & $\braket{\frac{1}{2}|0|\frac{1}{2}}_U$ & $\braket{1|0|1}_U$ \\
    \hline
    $\Sigma_c^+ \to p \gamma$ & $\sqrt{\frac{1}{3}}$ & $1$ & $0$\\
    $\Sigma_c^0 \to n \gamma$ & $\sqrt{\frac{2}{3}}$ & $0$ & $1$\\
    $\Xi_c^{\prime +} \to \Sigma^+ \gamma$ & $0$ & $1$ & $0$\\
    $\Xi_c^{\prime 0} \to \Lambda \gamma$ & $0$ & $0$ & $\frac{\sqrt{3}}{2}$\\
    $\Xi_c^{\prime 0} \to \Sigma^0 \gamma$ & $0$ & $0$ & $\frac{1}{2}$\\
    $\Omega_c \to \Xi^0 \gamma$ & $0$ & $0$ & $1$ 
  \end{tabular}
  \caption{Iso- and U-spin decomposition of the BSM contribution for the charm sextet baryons. The matrix elements are related to the amplitudes in Table \ref{tbl:Flavor Relationen sextet} as follows: $E_7 \sim \braket{\frac{1}{2}|0|\frac{1}{2}}_U$ and $E_7^\prime \sim \braket{1|0|1}_U$}
  \label{tbl:Iso- U-spin BSM sextet}
\end{table}
\FloatBarrier
\newpage
\section{$SU(3)_F$ decomposition of decay amplitudes}
\label{sec:SU(3) decomposition}

In this section we provide tables with the $SU(3)_F$ decomposition of the decay amplitudes for $B_{c3/6}\to B_8 \gamma$ and $B_{c6} \to B_{10}\gamma$. $SU(3)$ Clebsch Gordan coefficients can be determined using the isoscalar factors from \cite{Kaeding:1995vq}.

\begin{table}[h]
  \centering\begin{tabular}{l|c|c|c}
    Decay & $A_3$ & $A_{\overline{6}}$ & $A_{15}$ \\
    \hline
    $\Lambda_c \to \Sigma^+ \gamma$ & 0 & $\sqrt{\frac{2}{3}}V_{cs}^* V_{ud}$ & $-\frac{2}{\sqrt{15}}V_{cs}^* V_{ud}$ \\
    $\Xi_c^0 \to \Xi^0 \gamma$      & 0 & $\sqrt{\frac{2}{3}}V_{cs}^* V_{ud}$ & $\frac{2}{\sqrt{15}}V_{cs}^* V_{ud}$ \\
    \hline
    $\Lambda_c \to p \gamma$      & $\Delta$                   & $\sqrt{\frac{2}{3}}\Sigma$  & $-\frac{2}{\sqrt{15}} \Sigma -\frac{1}{\sqrt{15}} \Delta$ \\
    $\Xi_c^+ \to \Sigma^+ \gamma$ & $\Delta$                   & $-\sqrt{\frac{2}{3}}\Sigma$ & $\frac{2}{\sqrt{15}} \Sigma -\frac{1}{\sqrt{15}} \Delta$ \\
    $\Xi_c^0 \to \Lambda \gamma$  & $\sqrt{\frac{1}{6}}\Delta$ & $\Sigma$                    & $\sqrt{\frac{2}{5}} \Sigma +\frac{1}{\sqrt{10}} \Delta$ \\
    $\Xi_c^0 \to \Sigma^0 \gamma$ & $\sqrt{\frac{1}{2}}\Delta$ & $-\sqrt{\frac{1}{3}}\Sigma$ & $-\sqrt{\frac{2}{15}} \Sigma +\sqrt{\frac{3}{10}} \Delta$ \\
    \hline
    $\Xi_c^+ \to p \gamma$ & 0 & $\sqrt{\frac{2}{3}}V_{cd}^* V_{us}$ & $-\frac{2}{\sqrt{15}}V_{cd}^* V_{us}$ \\
    $\Xi_c^0 \to n \gamma$ & 0 & $\sqrt{\frac{2}{3}}V_{cd}^* V_{us}$ & $ \frac{2}{\sqrt{15}}V_{cd}^* V_{us}$ 
  \end{tabular}
  \caption{$SU(3)_F$ decomposition of the SM decay amplitudes for the charmed anti-triplet baryons.}
  \label{tbl:SU(3) triplet}
\end{table}

\begin{table}[h]
  \centering\begin{tabular}{l|c|c|c}
    Decay & $A^\prime_3$ & $A^\prime_{\overline{6}}$ & $A^\prime_{15}$ \\
    \hline
    $\Sigma_c^+ \to \Sigma^+ \gamma$    & 0 & $\sqrt{\frac{2}{5}}V_{cs}^* V_{ud}$   & $-\frac{2}{3\sqrt{5}}V_{cs}^* V_{ud}$ \\
    $\Sigma_c^0 \to \Lambda \gamma$     & 0 & $-\sqrt{\frac{2}{15}}V_{cs}^* V_{ud}$ & $-\sqrt{\frac{2}{15}}V_{cs}^* V_{ud}$ \\
    $\Sigma_c^0 \to \Sigma^0 \gamma$    & 0 & $\sqrt{\frac{2}{5}}V_{cs}^* V_{ud}$   & $-\frac{2}{3\sqrt{5}}V_{cs}^* V_{ud}$ \\
    $\Xi_c^{\prime 0} \to \Xi^0 \gamma$ & 0 & $\sqrt{\frac{2}{5}}V_{cs}^* V_{ud}$   & $\frac{2}{3\sqrt{5}}V_{cs}^* V_{ud}$ \\
    \hline
    $\Sigma_c^+ \to p \gamma$              & $-\sqrt{\frac{1}{3}}\Delta$ & $\sqrt{\frac{2}{5}}\Sigma$  & $-\frac{2}{3\sqrt{5}}\Sigma + \frac{1}{\sqrt{5}}\Delta$ \\
    $\Sigma_c^0 \to n \gamma$              & $-\sqrt{\frac{2}{3}}\Delta$ & $\sqrt{\frac{4}{5}}\Sigma$  & $\frac{2\sqrt{2}}{3\sqrt{5}}\Sigma - \frac{\sqrt{2}}{3\sqrt{5}}\Delta$ \\
    $\Xi_c^{\prime +} \to \Sigma^+ \gamma$ & $\sqrt{\frac{1}{3}}\Delta$  & $\sqrt{\frac{2}{5}}\Sigma$  & $-\frac{2}{3\sqrt{5}}\Sigma - \frac{1}{\sqrt{5}}\Delta$ \\
    $\Xi_c^{\prime 0} \to \Lambda \gamma$  & $-\sqrt{\frac{1}{2}}\Delta$ & $\sqrt{\frac{1}{15}}\Sigma$ & $-\sqrt{\frac{2}{15}}\Sigma - \frac{1}{\sqrt{30}}\Delta$ \\
    $\Xi_c^{\prime 0} \to \Sigma^0 \gamma$ & $\sqrt{\frac{1}{6}}\Delta$  & $\sqrt{\frac{1}{5}}\Sigma$  & $-\sqrt{\frac{2}{5}}\Sigma + \frac{1}{3\sqrt{10}}\Delta$ \\
    $\Omega_c \to \Xi^0 \gamma$            & $\sqrt{\frac{2}{3}}\Delta$  & $\sqrt{\frac{4}{5}}\Sigma$  & $\frac{2\sqrt{2}}{3\sqrt{5}}\Sigma + \frac{\sqrt{2}}{3\sqrt{5}}\Delta$ \\
    \hline
    $\Xi_c^{\prime +} \to p \gamma$ & 0 & $-\sqrt{\frac{2}{5}}V_{cd}^* V_{us}$          & $\frac{2}{3\sqrt{5}}V_{cd}^* V_{us}$ \\
    $\Xi_c^{\prime 0} \to n \gamma$ & 0 & $-\sqrt{\frac{2}{5}}V_{cd}^* V_{us}$          & $-\frac{2}{3\sqrt{5}}V_{cd}^* V_{us}$ \\
    $\Omega_c \to \Lambda \gamma$   & 0 & $-\frac{2\sqrt{2}}{\sqrt{15}}V_{cd}^* V_{us}$ & 0 \\
    $\Omega_c \to \Sigma^0 \gamma$  & 0 & 0                                             & $\frac{4}{3\sqrt{5}}V_{cd}^* V_{us}$  
  \end{tabular}
  \caption{$SU(3)_F$ decomposition of the SM decay amplitudes for the charmed sextet baryons.}
  \label{tbl:SU(3) sextet}
\end{table}
\FloatBarrier
\section{Flavor relations for $B_{c6} \to B_{10} \gamma$ }
\label{app:6to10}
We provide the flavor relations between the $B_{c6} \to B_{10} \gamma$ decays in Table \ref{tbl:Flavor Relationen decuplet}, as well as the U-spin and isospin decompositions in Table \ref{tbl:U-Spin decuplet} and \ref{tbl:Iso- U-spin BSM decuplet}, respectively. Note that the anti-sextet operator does not contribute to the $B_{c6} \to B_{10} \gamma$ amplitude, since the operator is symmetric under permutation of $u$ and $q^\prime$, while the flavor wave functions of the charmed sextet baryons and decuplet baryons are symmetric under permutation of quark flavors. Accordingly, the decomposition in irreducible representations of $\overline{6} \otimes 6 = 27 \oplus 8 \oplus 1$ does not include a decuplet, see Table \ref{tbl:SU(3) decuplet}.
\begin{table}[th]
  \centering\begin{tabular}{l|c|c|c}
    Decay  & U-Spin & $SU(3)_F$ & $SU(3)_F$ IRA\\
    \hline
    $\Sigma_c^+ \to \Sigma^{*+} \gamma$     & $V_{cs}^* V_{ud} H_\Sigma$        & $V_{cs}^* V_{ud} I$ & $V_{cs}^* V_{ud} J$\\
    $\Sigma_c^0 \to \Sigma^{*0} \gamma$     & $V_{cs}^* V_{ud} H^\prime_\Sigma$ & $V_{cs}^* V_{ud} I$ & $V_{cs}^* V_{ud} J$\\
    $\Xi_c^{\prime 0} \to \Xi^{*0} \gamma$  & $V_{cs}^* V_{ud} H^\prime_\Sigma$ & $V_{cs}^* V_{ud} I$ & $V_{cs}^* V_{ud} J$\\
    \hline
    $\Sigma_c^{++} \to \Delta^{++} \gamma$ & $\Delta H_\Delta^{\prime\prime} + H_7^{\prime \prime}$ & $-\sqrt{\frac{3}{2}} \Delta I + \sqrt{3} I_7$ & $-\sqrt{\frac{3}{2}} \Delta J + \sqrt{3} J_7$\\
    $\Sigma_c^{+} \to \Delta^{+} \gamma$   & $-\Sigma H_\Sigma + \Delta H_\Delta + H_7$ & $-\Sigma I + \sqrt{2} I_7$ & $-\Sigma J + \sqrt{2} J_7$\\
    $\Sigma_c^0 \to \Delta^0 \gamma$       & $-\sqrt{2} \Sigma H^\prime_\Sigma + \Delta H^\prime_\Delta + H_7^\prime$ & $-\sqrt{2}\Sigma I + \frac{1}{\sqrt{2}}\Delta I + J_7$ & $-\sqrt{2}\Sigma J + \frac{1}{\sqrt{2}}\Delta J + J_7$\\
    $\Xi_c^{\prime +} \to \Sigma^{*+} \gamma$  & $\Sigma H_\Sigma + \Delta H_\Delta + H_7$ & $\Sigma I + \sqrt{2}I_7$ & $\Sigma J + \sqrt{2}J_7$\\
    $\Xi_c^{\prime 0} \to \Sigma^{*0} \gamma$  & $\Delta H^\prime_\Delta + H_7^\prime$ & $\frac{1}{\sqrt{2}}\Delta I + I_7$ & $\frac{1}{\sqrt{2}}\Delta J + J_7$\\
    $\Omega_c \to \Xi^{*0} \gamma$  & $\sqrt{2} \Sigma H^\prime_\Sigma + \Delta H^\prime_\Delta + H_7^\prime$ & $\sqrt{2}\Sigma I + \frac{1}{\sqrt{2}}\Delta I + J_7$ & $\sqrt{2}\Sigma J + \frac{1}{\sqrt{2}}\Delta J + J_7$\\
    \hline
    $\Xi_c^{\prime +} \to \Delta^+ \gamma$  & $V_{cd}^* V_{us} H_\Sigma$        & $V_{cd}^* V_{us} I$ & $-V_{cd}^* V_{us} J$\\
    $\Xi_c^{\prime 0} \to \Delta^0 \gamma$  & $V_{cd}^* V_{us} H^\prime_\Sigma$ & $V_{cd}^* V_{us} I$ & $-V_{cd}^* V_{us} J$\\
    $\Omega_c \to \Sigma^{*0} \gamma$       & $V_{cd}^* V_{us} H^\prime_\Sigma$ & $V_{cd}^* V_{us} I$ & $-V_{cd}^* V_{us} J$
  \end{tabular}
  \caption{Flavor symmetry relations of the decay amplitudes for the charmed sextet baryons into decuplet baryons. Analogously to Table \ref{tbl:Flavor Relationen}, $H_\Sigma^{(\prime)}$, $H_\Delta^{(\prime, \prime \prime)}$ and $H_7^{(\prime, \prime \prime)}$ denote the U-spin triplet, U-spin singlet and the $c \to u \gamma$ contributions, respectively. Note that $\frac{1}{\sqrt{3}} H_7^{\prime \prime} = \frac{1}{\sqrt{2}} H_7^{\prime} = H_7$ in the isospin limit. Furthermore, $I= \frac{\sqrt{2}}{3}A^{\prime \prime}_{15}$ and $I_7 = \sqrt{\frac{1}{3}}(I_\text{NP} + \Delta A^{\prime \prime}_3)$. $J = \sqrt{\frac{2}{3}} \tilde{b}^{\prime\prime}_1 $ and  $J_7 = \frac{1}{\sqrt{3}}b_1^{\prime\prime}$ denote the weak annihilation and $c \to u \gamma$ contributions in the $SU(3)_F$ IRA.}
  \label{tbl:Flavor Relationen decuplet}
\end{table}
\begin{table}[th]
  \centering\begin{tabular}{l|c|c|c|c|c}
    Decay & $\braket{\frac{1}{2}|1|\frac{1}{2}}$ & $\braket{1|1|1}$  & $\braket{\frac{1}{2}|0|\frac{1}{2}}$ & $\braket{1|0|1}$ & $\braket{0|0|0}$\\
    \hline
    $\Sigma_c^+ \to \Sigma^{*+} \gamma$ & $\sqrt{\frac{2}{3}}V_{cs}^* V_{ud}$ & 0 & - & - & - \\
    $\Sigma_c^0 \to \Sigma^{*0} \gamma$ & 0 & $\frac{1}{\sqrt{2}}V_{cs}^* V_{ud}$ & - & - & - \\
    $\Xi_c^{\prime 0} \to \Xi^{*0} \gamma$ & 0 & $\frac{1}{\sqrt{2}}V_{cs}^* V_{ud}$ & - & - & - \\
    \hline
    $\Sigma_c^{++} \to \Delta^{++} \gamma$ & 0 & 0 & 0 & 0 & $\sqrt{2}\Delta$ \\
    $\Sigma_c^{+} \to \Delta^{+} \gamma$ & $-\sqrt{\frac{2}{3}}\Sigma$ & 0 & $\sqrt{2}\Delta$ & 0 & 0 \\
    $\Sigma_c^0 \to \Delta^0 \gamma$ & 0 & $-\Sigma$ & 0 & $\sqrt{2}\Delta$ & 0 \\
    $\Xi_c^{\prime +} \to \Sigma^{*+} \gamma$ & $\sqrt{\frac{2}{3}}\Sigma$ & 0 & $\sqrt{2}\Delta$ & 0 & 0 \\
    $\Xi_c^{\prime 0} \to \Sigma^{*0} \gamma$ & 0 & 0 & 0 & $\sqrt{2}\Delta$ & 0 \\
    $\Omega_c \to \Xi^{*0} \gamma$ & 0 & $\Sigma$ & 0 & $\sqrt{2}\Delta$ & 0 \\
    \hline
    $\Xi_c^{\prime +} \to \Delta^+ \gamma$ & $\sqrt{\frac{2}{3}}V_{cd}^* V_{us}$ & 0 & - & - & - \\
    $\Xi_c^{\prime 0} \to \Delta^0 \gamma$ & 0 & $\frac{1}{\sqrt{2}}V_{cd}^* V_{us}$ & - & - & - \\
    $\Omega_c \to \Sigma^{*0} \gamma$ & 0 & $\frac{1}{\sqrt{2}}V_{cd}^* V_{us}$ & - & - & - \\
  \end{tabular}
  \caption{U-Spin decomposition of the SM decay amplitudes for the decays of charmed sextet baryons into decuplet baryons.  In the matrix element $\braket{ U(f)| U(O) | U(i)}$, $U(f)$, $U(O)$ and $U(i)$ denote the U-spin of the final state, the U-spin changing operators and the initial state, respectively. The matrix elements are related to the amplitudes in Table \ref{tbl:Flavor Relationen decuplet} as follows: $H_\Sigma \sim \sqrt{\frac{2}{3}}\braket{\frac{1}{2}|1|\frac{1}{2}}$, $H_\Sigma^\prime \sim \frac{1}{\sqrt{2}}\braket{1|1|1}$, $H_\Delta \sim \sqrt{2}\braket{\frac{1}{2}|0|\frac{1}{2}}$, $H_\Delta^\prime \sim \sqrt{2}\braket{1|0|1}$ and $H_\Delta^{\prime\prime} \sim \sqrt{2}\braket{0|0|0}$.}
  \label{tbl:U-Spin decuplet}
\end{table}
\begin{table}[th]
  \centering\begin{tabular}{l|c|c|c|c|c}
    Decay & $\braket{\frac{3}{2}|\frac{1}{2}|1}_I$ & $\braket{1|\frac{1}{2}|\frac{1}{2}}_I$ & $\braket{\frac{1}{2}|0|\frac{1}{2}}_U$ & $\braket{1|0|1}_U$ & $\braket{0|0|0}_U$ \\
    \hline
    $\Sigma_c^{++} \to \Delta^{++} \gamma$    & $1$                  & 0                    & 0   & 0   & $1$ \\
    $\Sigma_c^{+} \to \Delta^{+} \gamma$      & $\sqrt{\frac{2}{3}}$ & 0                    & $1$ & 0   & 0 \\
    $\Sigma_c^0 \to \Delta^0 \gamma$          & $\sqrt{\frac{1}{3}}$ & 0                    & 0   & $1$ & 0 \\
    $\Xi_c^{\prime +} \to \Sigma^{*+} \gamma$ & 0                    & $1$                  & $1$ & 0   & 0 \\
    $\Xi_c^{\prime 0} \to \Sigma^{*0} \gamma$ & 0                    & $\sqrt{\frac{1}{3}}$ & 0   & $1$ & 0 \\
    $\Omega_c \to \Xi^{*0} \gamma$            & 0                    & 0                    & 0   & $1$ & 0 
  \end{tabular}
  \caption{Iso- and U-spin decomposition of the BSM contribution for decays of charm sextet baryons into decuplet baryons. The matrix elements are related to the amplitudes in Table \ref{tbl:Flavor Relationen decuplet} as follows: $H_7 \sim \braket{\frac{1}{2}|0|\frac{1}{2}}_U$, $H_7^\prime \sim \braket{1|0|1}_U$ and $H_7^{\prime \prime} \sim \braket{0|0|0}_U$}
  \label{tbl:Iso- U-spin BSM decuplet}
\end{table}
\FloatBarrier
\begin{table}[th]
  \centering\begin{tabular}{l|c|c|c}
    Decay & $A^{\prime \prime}_3$ & $A^{\prime \prime}_{\overline{6}}$ & $A^{\prime \prime}_{15}$ \\
    \hline
    $\Sigma_c^+ \to \Sigma^{*+} \gamma$    & 0 & 0 & $\frac{\sqrt{2}}{3}V_{cs}^* V_{ud}$ \\
    $\Sigma_c^0 \to \Sigma^{*0} \gamma$    & 0 & 0 & $\frac{\sqrt{2}}{3}V_{cs}^* V_{ud}$ \\
    $\Xi_c^{\prime 0} \to \Xi^{*0} \gamma$ & 0 & 0 & $\frac{\sqrt{2}}{3}V_{cs}^* V_{ud}$ \\
    \hline
    $\Sigma_c^{++} \to \Delta^{++} \gamma$    & $\Delta$                   & 0 & $-\sqrt{\frac{1}{3}} \Delta$ \\
    $\Sigma_c^{+} \to \Delta^{+} \gamma$      & $\sqrt{\frac{2}{3}}\Delta$ & 0 & $-\frac{\sqrt{2}}{3} \Sigma$ \\
    $\Sigma_c^0 \to \Delta^0 \gamma$          & $\sqrt{\frac{1}{3}}\Delta$ & 0 & $-\frac{2}{3} \Sigma + \frac{1}{3}\Delta$ \\
    $\Xi_c^{\prime +} \to \Sigma^{*+} \gamma$ & $\sqrt{\frac{2}{3}}\Delta$ & 0 & $\frac{\sqrt{2}}{3} \Sigma$ \\
    $\Xi_c^{\prime 0} \to \Sigma^{*0} \gamma$ & $\sqrt{\frac{1}{3}}\Delta$ & 0 & $\frac{1}{3} \Delta$ \\
    $\Omega_c \to \Xi^{*0} \gamma$            & $\sqrt{\frac{1}{3}}\Delta$ & 0 & $\frac{2}{3} \Sigma + \frac{1}{3}\Delta$ \\
    \hline
    $\Xi_c^{\prime +} \to \Delta^+ \gamma$ & 0 & 0 & $\frac{\sqrt{2}}{3}V_{cs}^* V_{ud}$ \\
    $\Xi_c^{\prime 0} \to \Delta^0 \gamma$ & 0 & 0 & $\frac{\sqrt{2}}{3}V_{cs}^* V_{ud}$ \\
    $\Omega_c \to \Sigma^{*0} \gamma$      & 0 & 0 & $\frac{\sqrt{2}}{3}V_{cs}^* V_{ud}$ 
  \end{tabular}
  \caption{$SU(3)_F$ decomposition of the SM decay amplitudes for the decays of charmed sextet baryons into decuplet baryons.}
  \label{tbl:SU(3) decuplet}
\end{table}
\section{$SU(3)_F$ irreducible representation approach amplitudes \label{app:SU3}}
The $SU(3)_F$ irreducible representation approach  amplitudes are defined by \cite{Wang:2020wxn}
\begin{align}
  \begin{split}
    \A(B_{c3} \to B_8 \gamma) &= b_1 (B_{c3})^{[ij]}T^\prime(\overline{3})^k (B_8)_{[ij]k} + b_2 (B_{c3})^{[ij]}T^\prime(\overline{3})^k (B_8)_{j[ik]} \\
                              &+ \left(\tilde{b}_1 H(\overline{6})_j^{lk} + \tilde{b}_4 H(15)_j^{lk}\right) (B_{c3})^{[ij]}(B_8)_{k[il]} \\
                              &+ \left(\tilde{b}_2 H(\overline{6})_j^{lk} + \tilde{b}_5 H(15)_j^{lk}\right) (B_{c3})^{[ij]}(B_8)_{l[ik]} \\
                              &+ \left(\tilde{b}_3 H(\overline{6})_j^{lk} + \tilde{b}_6 H(15)_j^{lk}\right) (B_{c3})^{[ij]}(B_8)_{i[lk]}\, , 
  \end{split}
\end{align}
\begin{align}
  \begin{split}
    \A(B_{c6} \to B_8 \gamma) &= b_1^\prime (B_{c6})^{ij}T^\prime(\overline{3})^k (B_8)_{j[ik]} \\
                              &+ \left(\tilde{b}_1^\prime H(\overline{6})_j^{lk} + \tilde{b}_4^\prime H(15)_j^{lk}\right) (B_{c6})^{ij}(B_8)_{k[il]} \\
                              &+ \left(\tilde{b}_2^\prime H(\overline{6})_j^{lk} + \tilde{b}_5^\prime H(15)_j^{lk}\right) (B_{c6})^{ij}(B_8)_{l[ik]} \\
                              &+ \left(\tilde{b}_3^\prime H(\overline{6})_j^{lk} + \tilde{b}_6^\prime H(15)_j^{lk}\right) (B_{c6})^{ij}(B_8)_{i[lk]}\, , 
  \end{split}
\end{align}
\begin{align}
    \A(B_{c6} \to B_{10} \gamma) = b_1^{\prime \prime} (B_{c6})^{ij}T^\prime(\overline{3})^k (B_{10})_{ijk}  + \tilde{b}_1^{\prime\prime} H(15)_j^{lk} (B_{c6})^{ij}(B_{10})_{ilk} \, , 
\end{align}
where $T^\prime(\overline{3}) = (1,0,0)$, $(B_{c3})^{[ij]} = \epsilon^{ijk} (B_{c3})_k$ and $(B_8)_{i[jk]} = \epsilon_{jkx} (B_8)^{~x}_{i}$. The charm baryon anti-triplet $B_{c3}$, sextet $B_{c6}$, the light baryon octet $B_8$ and decuplet $B_{10}$ can be written as
\begin{align}
    &B_{c3}=(\Xi_c^0, -\Xi_c^+, \Lambda_c)\, , \quad B_{c6} = \begin{pmatrix}
      \Sigma_c^{++} & \frac{1}{\sqrt{2}}\Sigma_c^+ & \frac{1}{\sqrt{2}}\Xi_c^{\prime +} \\
      \frac{1}{\sqrt{2}} \Sigma_c^+ & \Sigma_c^0 & \frac{1}{\sqrt{2}}\Xi_c^{\prime 0} \\
      \frac{1}{\sqrt{2}} \Xi_c^{\prime +} & \frac{1}{\sqrt{2}}\Xi_c^{\prime 0} & \Omega_c\end{pmatrix} \, , \quad B_8 = \begin{pmatrix}
        \frac{\Lambda}{\sqrt{6}} + \frac{\Sigma^0}{\sqrt{2}} & \Sigma^+ & p \\
        \Sigma^- & \frac{\Lambda}{\sqrt{6}} - \frac{\Sigma^0}{\sqrt{2}} & n \\
        \Xi^- & \Xi^0 & -\frac{2\Lambda}{\sqrt{6}} 
    \end{pmatrix}\,,\nonumber\\
    &B_{10} = \frac{1}{\sqrt{3}} \begin{pmatrix}
      \begin{pmatrix}
        \sqrt{3}\Delta ^{++} & \Delta^+ & \Sigma^{*+} \\
        \Delta^+ & \Delta^0 & \frac{\Sigma^{*0}}{\sqrt{2}} \\
        \Sigma^{*+} & \frac{\Sigma^{*0}}{\sqrt{2}} & \Xi^{*0}
      \end{pmatrix}, \begin{pmatrix}
        \Delta^+ & \Delta^0 & \frac{\Sigma^{*0}}{\sqrt{2}} \\
        \Delta^0 & \sqrt{3} \Delta^- & \Sigma^{*-} \\
        \frac{\Sigma^{*0}}{\sqrt{2}} & \Sigma^{*-} & \Xi^{*-}
      \end{pmatrix}, \begin{pmatrix}
        \Sigma^{*+} & \frac{\Sigma^{*0}}{\sqrt{2}} & \Xi^{*0} \\
        \frac{\Sigma^{*0}}{\sqrt{2}} & \Sigma^{*-} & \Xi^{*-} \\
        \Xi^{*0} & \Xi^{*-} & \sqrt{3}\Omega^- 
      \end{pmatrix}
    \end{pmatrix}
\end{align}
Due to the (anti)symmetry of ($H(\overline{6})_j^{lk}$) $H(15)_j^{lk}$ in $l$ and $k$, the following relations hold
\begin{equation}
  \tilde{b}^{(\prime)}_2 = - \tilde{b}^{(\prime)}_1\, , \quad \tilde{b}^{(\prime)}_4 = \tilde{b}^{(\prime)}_5\, , \quad \tilde{b}^{(\prime)}_6=0\, .
\end{equation}
The element of the tensors $H(\overline{6})$ and $H(15)$ are given in \cite{Wang:2017azm}. For Cabibbo favored $c \to su \dbar$ transitions the non-zero elements are
\begin{align}
  H(\overline{6})_2^{31} = - H(\overline{6})_2^{13} = 1\, , \qquad H(15)_2^{31} = H(15)_2^{13} = 1\, ,
\end{align}
with an overall factor of $V_{cs}^* V_{ud}$. For doubly Cabibbo suppressed $c \to du \sbar$ transitions the non-zero elements are
\begin{align}
  -H(\overline{6})_3^{21} = H(\overline{6})_3^{12} = 1\, , \qquad H(15)_3^{21} = H(15)_3^{12} = -1\, ,
\end{align}
with an overall factor of $V_{cd}^* V_{us}$. For singly Cabibbo suppressed $c \to u d \dbar$ transitions
\begin{align}
  \begin{split}
    &H(\overline{6})_2^{21} = - H(\overline{6})_2^{12} = H(\overline{6})_3^{13} = - H(\overline{6})_3^{31} = \frac{1}{2}\, ,\\
    &\frac{1}{3}H(15)_2^{21} = \frac{1}{3}H(15)_2^{12} = -\frac{1}{2}H(15)_1^{11} = -H(15)_3^{13} = -H(15)_3^{31} = \frac{1}{4}\, ,
  \end{split}
\end{align}
with an overall factor of $V_{cd}^* V_{ud}$. For singly Cabibbo suppressed $c \to u s \sbar$ transitions
\begin{align}
  \begin{split}
    &-H(\overline{6})_2^{21} = H(\overline{6})_2^{12} = -H(\overline{6})_3^{13} = H(\overline{6})_3^{31} = \frac{1}{2}\, ,\\
    &-H(15)_2^{21} = -H(15)_2^{12} = -\frac{1}{2}H(15)_1^{11} = \frac{1}{3}H(15)_3^{13} = \frac{1}{3}H(15)_3^{31} = \frac{1}{4}\, ,
  \end{split}
\end{align}
with an overall factor of $V_{cs}^* V_{us}$.

\section{$B_c \to B$ form factors}
\label{app:form_factors}

The $B_c \to B$  tensor form factors can be written as
\begin{align}
  \begin{split}
  \langle B(q,\,s_B) \vert \overline{u} \text{i} \sigma^{\mu\nu}k_\nu c\vert B_c(P,\,s_{B_c}) \rangle =&\\
  -\overline{u}(q,\,s_B)& \left[\frac{h_+^{B_c \to B}(k^2)}{s_+}k_\nu\left(k^\nu s^\mu - s^\nu k^\mu\right)\right.\\
  &\left. + h_\perp^{B_c \to B}(k^2)\left(-\im \sigma^{\mu\nu} k_\nu  + \frac{1}{s_+} k_\nu\left(s^\nu k^\mu - k^\nu s^\mu\right)\right)\right] u(P,\,s_{B_c}) \, ,
  \end{split}
\end{align}
\begin{align}
  \begin{split}
  \langle B(q,\,s_B) \vert \overline{u} \text{i} \sigma^{\mu\nu}k_\nu \gamma_5 c\vert B_c(P,\,s_{B_c}) \rangle =&\\
  -\overline{u}(q,\,s_B)\gamma_5& \left[\frac{\tilde{h}_+^{B_c \to B}(k^2)}{s_-}k_\nu\left(k^\nu s^\mu - s^\nu k^\mu\right)\right.\\
  &\left. + \tilde{h}_\perp^{B_c \to B}(k^2)\left(-\im \sigma^{\mu\nu} k_\nu  + \frac{1}{s_-} k_\nu\left(s^\nu k^\mu - k^\nu s^\mu\right)\right)\right] u(P,\,s_{B_c}) \, ,
  \end{split}
\end{align}
with $k=P-q$, $s=P+q$ and $s_{\pm}=(m_{B_c}\pm m_B)^2-k^2$. The definition of the tensor form factors is  identical to those in \cite{Meinel:2017ggx}, however, 
above we rearranged the kinematic quantities in a way which is more practical  for radiative decays. To  the latter only $h_\perp$ and $\tilde h_\perp$ contribute due to gauge invariance. Moreover, $h_\perp(k^2=0)=\tilde h_\perp(k^2=0)$ exactly~\cite{Hiller:2021zth}. To estimate the NP reach we employ results from lattice QCD \cite{Meinel:2017ggx},
$h_\perp(0)^{\Lambda_c \to p}=0.511 \pm 0.027$ and flavor symmetries (\ref{eq:ff3to8}).


\end{document}